\documentclass[nofootinbib,eqsecnum,tightenlines,11pt]{revtex4}

\usepackage{graphicx}
\usepackage{amsmath,amssymb,amsfonts,amsthm,stmaryrd,mathtools}
\usepackage{mathrsfs}
\usepackage{color}
\usepackage{multirow,bigdelim}
\usepackage{dsfont}
\allowdisplaybreaks[0]

\usepackage{axodraw4j}
\usepackage{pstricks}

\newcommand{\pic}[1]{
\vcenter{\hbox{\includegraphics[scale=0.6]{#1}}}
}

\def\be#1\ee{\begin{align}#1\end{align}}

\def\ba{\begin{eqnarray}}
\def\ea{\end{eqnarray}}

\def\f{\frac}

\def\nn{\nonumber}
\def\q{\qquad}
\def\hs{\hspace{-0.3cm}}

\def\i{\mathrm{i}}
\def\SU{{\rm{SU}}}

\def\SO{\mathrm{SO}}

\def\su{\mathfrak{su}}

\def\D{\mathcal{D}}

\def\k{{\rm k}}
\def\tk{{\tilde{\rm k}}}
\def\D{D_{\rm T}}
\begin{document}

\title{Cosmological constant from  condensation of defect excitations}

\author{Bianca Dittrich}
\affiliation{Perimeter Institute for Theoretical Physics,\\ 31 Caroline Street North, Waterloo, Ontario, Canada N2L 2Y5}

\begin{abstract}

A key challenge for many quantum gravity approaches is to construct states that describe smooth geometries on large scales. Here we  define a family of  $(2+1)$--dimensional quantum gravity states  which arise from curvature excitations concentrated at point like defects and describe homogeneously curved geometries on large scales. These states represent therefore vacua for three--dimensional gravity with different values of the cosmological constant. They can be described by an anomaly--free first class constraint algebra  quantized on one and the same Hilbert space for different values of the cosmological constant.   A similar construction is possible in four  dimensions, in this case the curvature is concentrated along string--like defects and the states are vacua  of the Crane--Yetter model. 
We will sketch applications for quantum cosmology and condensed matter.

\end{abstract}

\maketitle

\section{Introduction}

Three--dimensional gravity with Euclidean signature and positive cosmological constant can be described by the Turaev--Viro partition function \cite{TV} with a quantum deformed structure group $\SU(2)_\k$.  The Turaev--Viro models are defined for  triangulated three--dimensional manifolds, the partition function is however invariant under changes of the (bulk) triangulation.  Here $\k \in \mathbb{N}$ is the so--called level which determines the quantum deformation parameter $q=\exp(2\pi {\rm i}/(\k+2))$. The value of the cosmological constant $\Lambda$ is  encoded in the level $\k$
\ba
\k \,=\, \frac{1}{G\hbar \sqrt{\Lambda}}
\ea
with $G$ the three--dimensional Newton's constant.  We thus have a quantized cosmological constant $\Lambda = 1/(G^2\hbar^2 \k^2)$.

An earlier variant of the Turaev--Viro model is the Ponzano--Regge model \cite{PR}, which describes Euclidean three--dimensional gravity without a cosmological constant. The Ponzano--Regge model can be also understood as a zero--coupling limit of three--dimensional lattice gauge theory with $\SU(2)$ structure group.

That these models do in fact describe gravity can be shown by a semi-classical analysis of their amplitudes \cite{MiziguchiTada}. This analysis reproduces the Regge action, which describes discretized  gravity based on flat \cite{Regge}  or homogeneously curved tetrahedra \cite{Improved,NewRegge}.  Furthermore, both the Ponzano--Regge and Turaev--Viro (henceforth TV) model are topological theories, as is three--dimensional gravity. 
This is also the case  for  three--dimensional Regge gravity, if one uses homogeneously curved tetrahedra for the theory with cosmological constant \cite{Improved}. 

The dynamics of the Turaev--Viro model can be also described in a canonical form, e.g. with loop quantum gravity techniques \cite{SmolinLambda}  or within the string net formalism \cite{LevinWen,KKR,Kir,DG16}. By choosing a $\Lambda$ adapted vacuum state for the loop quantum gravity Hilbert space -- instead of a vacuum peaked on flat geometries \cite{DG14a} or a totally degenerate geometry \cite{AL} --  one can bring these two formulations together \cite{DG16}. 

In the canonical formalism one distinguishes between the kinematics and dynamics. Now in the works cited above the cosmological constant is already built into the kinematics, simplifying indeed enormously the imposition of the dynamics. 

In this paper we show that one can separate kinematics and dynamics, that is express the dynamics and solutions for different cosmological constants in the same kinematical Hilbert space. This does in particular allow to express solutions for different cosmological constant, which are peaked on different values of the homogeneous curvature, in the same Hilbert space. A solution with a cosmological constant $\tilde \Lambda>\Lambda$ can then be analyzed with regard to the curvature excitations on top of a $\Lambda$--vacuum. These are excitations which are also induced by massive (here spin-less) point particles. The $\tilde \Lambda$ vacuum can thus be understood to arise from a very particular superposition of such excited states.

This work opens up a number of further research directions and applications in quantum gravity and condensed matter:

~\\
{\bf Quantum gravity:} 
This work does pick up on a research direction to derive the quantum group deformation from the imposition of the cosmological constant \cite{PerezPranzetti,Improved,Etera17}. One main difficulty here is, that as one is  involving a discretization to define a diffeomorphism invariant dynamics, one encounters generically anomalies in the diffeomorphism constraint algebra \cite{PerezPranzetti} or covariantly, breaks diffeomorphism symmetry \cite{diffeobreak}. Here we will present an anomaly free quantization of the constraints defining the $\tilde \Lambda$ dynamics (for one of the allowed discrete values) in a kinematical Hilbert space associated to any allowed $\Lambda<\tilde \Lambda$.  

An alternative way to construct such quantizations which preserve diffeomorphism symmetry is to subject a given, not necessarily diffeomorphism invariant discrete theory, to a coarse graining flow. Each coarse graining step constructs an effective dynamics for a coarser lattice which reflects the dynamics of the finer lattice. Iterating the procedure one is performing a refinement limit and thus reaches a fixed point of the coarse graining flow. One can then hope that diffeomorphism symmetry is restored in this limit \cite{Improved,CGReviews,Hypercube}.

This strategy has been proposed in \cite{Improved} and has also been demonstrated to work for classical Regge calculus with a cosmological constant. It would be of course very useful to see that this can be extended to the quantum level, and this does indeed seem to be in reach \cite{BDtoappear}. 

~\\
{\bf Quantum groups in quantum gravity:} 
As mentioned above it would be highly interesting to derive a quantum group structure, and in particularly the braiding in a quantum group, by imposing the dynamics of 3D gravity with a cosmological constant. To this end one uses a first order formulation of 3D gravity with triads $e$ and a ($\su(2)$ valued) spin connection $A$ . The constraints  are then given by the curvature constraints
$
F-\Lambda e\wedge e =0  
$
where $F$ is the curvature of the spin connection. One also has the Gau\ss~constraint (or no--torsion condition) $d_Ae=0$. The curvature constraints are relatively simple to impose for $\Lambda=0$ as in this case the constraints only involve the connection and one can work in a connection polarization. For non--vanishing $\Lambda$ however the constraints involve canonically conjugated variables.  The constraints can be however rewritten into $C=\tfrac{1}{2}(F_++F_-)$ and $G=\tfrac{1}{2}(F_+-F_-)$ for the curvature and Gau\ss~constraints respectively. This uses the curvatures $F_\pm$ of two Poincare connections $A_{\pm}(\Lambda)=A\pm \sqrt{\Lambda} e$. The price to pay is that the components of the redefined connections become non--commutative. 

For the quantization of the theory one works with holonomies or Wilson lines. Noui, Perez and Pranzetti  \cite{NouiEtal} constructed the holonomy for $A_{\pm}(\Lambda)$ in terms of the holonomy of the spin connection and triad operators. The non--commutativity of the connection $A_\pm$ does result in a braiding structure for the Wilson lines that reproduces thus of the quantum group deformation $\SU(2)_\k$ \cite{NouiEtal}. It was pointed out however that the Wilson loops do not reproduce the quantum dimension due to the fact that one does not evaluate these on a solution of the constraints (which have not been constructed in \cite{NouiEtal}). 

Here, to have to deal with only one kinematical framework, we will rather work  with the case of having two different cosmological constants $\tilde \Lambda <\Lambda$.  Apart from the constraints we will  express the Wilson loop operators (which are here ribbon operators describing both the $A_+$ and the $A_-$ connection)  associated to $A_\pm(\tilde \Lambda)$ within the kinematical Hilbert space based on $\Lambda$. By construction these Wilson loop operators will reproduce the braiding associated to the level $\tk = 1/(G\hbar \sqrt{\tilde\Lambda})$. These Wilson loop operators also do reproduce the $\tk$--quantum dimensions, but only if considered on $\tk$--vacuum states.

~\\
{\bf Quantum cosmology:}  One strategy for deriving a cosmological dynamics from quantum gravity is to consider states which are in a certain sense homogeneous. This approach has been taking in loop quantum cosmology \cite{Bojowald} based on a notion of homogeneous connection \cite{BojowaldSR}. Although there are a number of works which made this notion more precise in the quantum realm \cite{HomConn}, the construction of loop quantum cosmology does rather start with a symmetry reduced classical phase space and does not consider a notion of quantum homogeneous states inside the full theory.\footnote{A first step in this direction was taken in \cite{AlesciC}, where the homogeneity is based on conditions on the triads. The eigenvalues of the triads characterize the size of a fundamental cell and this can be seen as a discretization ambiguity, see also the discussion in \cite{AlesciDensityMatrix}. An alternative framework is group field theory cosmology \cite{GFTcosmo}. Group field theory is based on similar constructions as loop quantum gravity \cite{OritiLQG} but in a certain sense includes a sum over discretizations. }

Here we construct a family of states peaked on homogeneous curvature. These states enjoy a certain discretization independence and therefore also diffeomorphism symmetry. Let us note that the construction, which we present here for the three--dimensional case, can be also done for the four--dimensional theory. In this case we can express the vacua state of the (topological) Crane--Yetter model \cite{CraneYetter} for different levels $\tk$ using the same kinematical Hilbert space based on the Crane--Yetter vacuum with some larger $\k$--level. This kinematical Hilbert space has been recently constructed in \cite{BD17} and can be understood as a quantum deformation of the Hilbert space for loop quantum gravity based on a flat curvature vacuum \cite{DG14a}.  Thus a notion of homogeneous states is also available in $(3+1)$ dimensions. 

A related proposal for the $(3+1)$ dimensional case has been recently put forward in \cite{Engle}, but employing a complexified (Poincare like) connection. A detailed quantization of the constraints defining the homogeneous states  is not yet available and  solutions have not been constructed. In our work we present a simple strategy  to construct a consistent set of homogeneity constraints on the quantum level and also the solutions to these constraints.  These steps can be generalized in a straightforward way to $(3+1)$ dimensions \cite{BD17}. We expect furthermore that the constructions can be generalized to Lorentzian signature which would imply a complexified connection. 

Having a family of homogeneous states $\psi_\tk$  available one can ask for a superposition $\phi$ of such states which would satisfy
\ba
\langle \psi_\tk \, | \, H \, | \, \phi\rangle \,= 0 \q ,
\ea
where $H$ is the Hamiltonian constraint with constant lapse. The solution $\phi$ would represent an approximation to the solution of the full theory, in the sense that we consider only a subset of the conditions that a full solution would have to satisfy.  

Allowing different patches or domains of the spatial hypersurface to carry different values of the homogeneous curvature we would need to consider domain walls, which we will discuss in the next point. Such domain walls will also appear in time direction if there is a non--trivial dynamics for the constant curvature value. 

Going back to the $(2+1)$--dimensional theory we would need to include a matter coupling to have a non--trivial dynamics. One would thus also need a homogeneity condition for the matter field. Such a set--up can  be used to study a matter--induced change of the cosmological constant. 

Here we have discrete values for the cosmological constant -- changes in its value could therefore show up as phase transitions. Such phase transitions are also of interest in condensed matter.

~\\
{\bf Condensed matter:}  
Being able to separate kinematics and dynamics we can define $\tk$--Hamiltonians ${\bf H}_\tk$ whose ground states are given by the $\tk$--vacua. We can thus consider linear combinations of two such Hamiltonians ${\bf H}(\alpha)=\alpha {\bf H}_\k +(1-\alpha) {\bf H}_\tk$. Varying the coupling constant  $\alpha$ we will obtain quantum phase transitions between the two phases characterized by $\tk$ and $\k$. That is we expect to see that in the limit of large lattices the gap between the energies for the ground state(s) and the first excited state(s) vanishes for a certain  $\alpha$. A phase transition has to occur as the degeneracy for non--trivial topologies changes if we change the level, e.g. for the torus we have a degeneracy $(\k+1)^2$ and $(\tk+1)^2$ respectively. We will furthermore define order parameter which indicate such transitions by taking on a vanishing value in one phase and a non--vanishing value in the other phase.

If these phase transitions are of second or higher order we expect that the corresponding phase transition points define a  $(2+1)$--dimensional conformal field theory with propagating degrees of freedom. There does not seem to be known much yet about these transitions and their order, but as there are many of these transitions they could turn out to be a rich source of such conformal field theories. 

There is another process that leads to phases for such $\SU(2)_\k$ systems as considered here, known as anyon condensation \cite{Bais}. The transitions between $\k$ and $\tk$ are however (generically) different from anyon condensation. In the case of anyon condensation one has a `sharply' defined subset of anyonic excitations (that is curvature and torsion excitations characterized by certain eigenvalues of the Wilson loop operators) that condense into a new vacuum state. Here we can analyze the $\tk$  vacuum with respect to the $\k$ vacuum and it will (in general) include a superposition of all (purely magnetic) anyonic excitations.  This said we can of course also consider such anyon condensation processes for varying levels $\tk$.

As the set--up is generalizable to $(3+1)$ dimensions \cite{WW,BD17}, we can expect similar phase transitions there. Although the ground states are unique even for non--trivial topology one has also in this case an order parameter that distinguishes the $\k$ and $\tk$ phases.

One can also realize different phases in different regions or domains (of the spatial hypersurface). In this case the domain wall, which separates two phases, consists at least of a row of plaquettes. The reason is that the plaquette projector  which make up the Hamiltonian, do not commute for different levels $\k$. There are thus degrees of freedom associated to the domain wall. 

Such a set--up could be also interesting for quantum cosmology. Firstly a domain wall, if seen as transversal to a time direction would describe a (time) transition between states peaked on different  homogeneous curvature values. Secondly patches with different homogeneous curvature can be used as a way to introduce inhomogeneities.

~\\
{\bf Lattice gauge theory and tensor networks:}  The kinematical Hilbert space we construct can be also interpreted as a Hilbert space for $(2+1)$-dimensional lattice gauge theory, where the structure group is  $q$--deformed to $\SU(2)_\k$. The $\tk$ vacua can in this sense also be understood as phases of lattice gauge theory, which are characterized by non--vanishing expectations values of the curvature. 

The $q$--deformation at the root of unity provides furthermore a cut--off, in the sense that it reduces the infinite--dimensional Hilbert space for $\SU(2)$ to a finite--dimensional one. This is important for numerical coarse graining techniques, such as tensor network algorithms  for lattice gauge theories \cite{DittTNWGauge, Milsted}. The finiteness of such $q$--deformed models has also been used to investigate certain two--dimensional models, designed to reproduce key dynamical mechanisms of spin foams \cite{spinnet}.  The fusion basis \cite{KKR,ABC16a}, which we will make also use of here, is also a useful tool for canonical and covariant coarse graining schemes \cite{DittSteintoappear}.

From a lattice gauge theory perspective one is in particular interested in the transition from $\tk=0$ to some large $\k$. The $\tk=0$ vacuum coincides with the strong coupling limit of lattice gauge theory, the $\k$--vacuum for $\k$ large represents a small deformation of the weak coupling limit. One can thus study the transition from the weak to the strong coupling regime as a function of $\k$. This has the advantage of working with manifestly finite Hilbert spaces for finite $\k$. As the dynamics is also adjusted to the $\k$--level it presents a far more elegant cut--off than truncating away all representation labels larger than some $j_{\text{max}}$ \cite{Milsted}.

~\\
{\bf Outline of the paper:} In the next section \ref{covariant} we will give a definition of the Turaev--Viro (TV) model and a related family of models. We will then follow up with a Hamiltonian description in section \ref{canonical}. We will first introduce the string net framework that defines a physical Hilbert space associated to the TV model. We then define a kinematical Hilbert space by introducing punctures that can carry defect excitations and thus allow for a dynamics different from the TV model.  The kinematical Hilbert spaces still carry a $\k$--dependent structure, but we will define an embedding of the $\tk$--Hilbert space into the $\k$--Hilbert space for $\tk<\k$. Using the simplest example, namely the two--punctured sphere, we then give a detailed description of the framework in section \ref{2punct}. In section \ref{general} we will consider general triangulations and show how to construct the various Hamiltonians, ribbon operators and $\k$-- and $\tk$--vacua in this case. This will also allow us to analyze the $\tk$--vacua in terms of the $\k$ excitations. We close with a discussion and outlook in section \ref{discussion}. The appendices \ref{SU2Ess}  to \ref{TTetra} collect various more technical background material, in particular some essential definitions related to $\SU(2)_\k$.

\section{The Turaev--Viro partition function and related models}\label{covariant}

The TV model  based on $\SU(2)_\k$ defines a partition function for 3D triangulations. The partition functions sums tetrahedral weights over  labels associated to edges. These labels are given by $\SU(2)_\k$ irreducible representations (irreps) $j_e \in \{0,\tfrac{1}{2},1,\ldots \tfrac{\k}{2}\}$:
\ba\label{TV1}
{\cal Z}_\k &=& \frac{1}{\cal N} \, \sum_{\{j_e\}}  \prod_e v^2_{j_e}  \prod_\tau {\cal A}_\tau( \{j_e\})  \q .
\ea
Here the amplitudes associated to the tetrahedra are given by
\ba\label{TV2}
{\cal A}_\tau( \{j_e\}) = \prod_{e \subset \tau}   \, G(\{j_e\}) 
\ea
and $G(j_1,\ldots,j_6)$ is (one version of) the $6j$ recoupling symbol. (See appendix \ref{SU2Ess}.) The six irreps $j_e$ are associated to the six edges of the tetrahedron as shown in the following figure: 
\ba
\pic{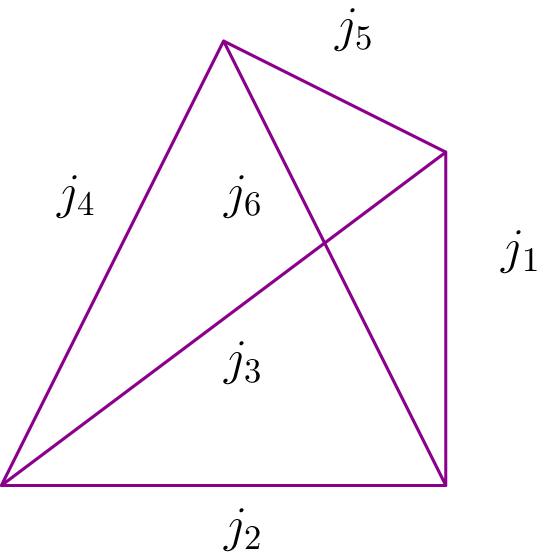} \nn \q .
\ea
  The $v_j^2=(-1)^{2j}d_j$ are the signed quantum dimension associated to the irreps $j$ and appear in (\ref{TV1}) as measure factors. Furthermore the normalization ${\cal N}$ in (\ref{TV1}) depends only on the number of vertices
${\cal N}\,=\,{\D}^{2\sharp {\rm v}}$
where ${\D}$ is the total quantum dimension of $\SU(2)_\k$ and $\sharp {\rm v}$ is the number of vertices in the triangulation.

 Note that as the tetrahedral amplitude is being given by an $\SU(2)_\k$ recoupling symbol it will be only non--vanishing if the following coupling conditions are satisfied for any of the triples $(j_e,j_{e'},j_{e''})_{e,e',e'' \in t}$ of irreps labelling a triangle:
\ba\label{Admissibility}
j_e\leq j_{e'}+j_{e''},
\;\;\;\;
j_{e'}\leq  j_{e}+j_{e''},
 \;\;\;\;
j_{e''}\leq  j_{e}+j_{e'},
\;\;\;\;
j_e+ j_{e'}+j_{e''}\in\mathbb{N},
\;\;\;\;
 j_e+ j_{e'}+j_{e''}\leq{\rm k} \; .
\ea

The TV partition function can also be defined for a triangulation with  boundary -- in this case the sum is only over the $j$'s associated to bulk edges.  For boundary edges one uses a measure factor $v_{j_e}$ (instead of $v_{j_e}^2$) and the normalization changes to
\ba
{\cal N}\,=\, {\cal D}^{-2 \sharp_{\circ} {\rm v} -\sharp_\partial {\rm v} }
\ea
with $\sharp_{\circ} {\rm v}$ the number of bulk vertices and $\sharp_\partial {\rm v}$ the number of vertices in the boundary.

~\\

The partition function (\ref{TV1}) is invariant under changes of the bulk triangulation and thus defines for closed manifolds a topological invariant. From this bulk invariance it also follows that for manifolds with boundary which are homomorphic to $\Sigma \times [0,1]$  the partition function can be understood to act as a projector on the so--called {\it kinematical} boundary Hilbert space, which we will define in section \ref{kinHil}. 

Thus the partition function (\ref{TV1}) has very special properties. We can consider this partition function as a very particular example of a family of models, which are of the same form as (\ref{TV1}), e.g. can be written as
\ba\label{TV3}
{\cal Z} &=& \frac{1}{{\cal N}'} \, \sum_{\{j_e\}}  \prod_e  (v'(j_e))^2 \prod_\tau {\cal A}'_\tau( \{j_e\}) 
\ea
with some normalization factor ${\cal N}'$, possibly depending on the number of vertices, tetrahedral amplitudes ${\cal A}'_\tau(\{j_e\})$, which are functions of $\SU(2)_\k$ irreps assigned to the edges of this tetrahedron and weight factors $(v'(j_e))$ associated to the edges. Here we will assume that the amplitude ${\cal A}'_\tau$ vanishes if the coupling conditions (\ref{Admissibility}) are not satisfied. This will allow us to express these amplitudes as states in the kinematical boundary Hilbert space associated to the $\SU(2)_\k$ models. 

We can now consider a renormalization flow of partition functions of the type (\ref{TV3}). That is we glue the tetrahedra  (or more general building blocks) to larger building blocks, integrate out the resulting bulk variables and obtain effective amplitudes for these new building blocks. These effective amplitudes will depend on a larger set of boundary data than the initial ones. For this reason one includes a truncation procedure: the boundary data are separated into relevant and irrelevant sets according to some criteria and only the relevant part is kept.  This allows one to consider the coarse graining flow  in a fixed space  of amplitudes, that is functions, that depend on a certain fixed number of variables that encode the allowed boundary data.  

Such coarse graining flows can be realized via tensor network algorithms \cite{TNWAlg}, whose various incarnations differ in particular in the criteria that separate relevant and irrelevant data.  See in particular \cite{spinnet} for algorithms involving 2D models with a quantum group symmetry and \cite{DittTNWGauge} for algorithms for 3D (generalized) lattice gauge models. Let us note that for non-Abelian lattice gauge models as well as the models considered here one has to generally expect that the flow does not preserve the coupling conditions \eqref{Admissibility}, see \cite{EteraCoarse,DG14a}. The crucial point here is the truncation step -- this can be actually chosen such that is always does project back onto amplitudes satisfying the coupling conditions \cite{DittTNWGauge}. However whether this only discards irrelevant data or not depends on the dynamics, i.e. the amplitudes, of the initial model. For gravitational models  one might be in particular concerned if the dynamics described in the initial model is expected to describe  non-vanishing homogeneous curvature -- as curvature leads to torsion excitations that lead to the violation of the coupling conditions.   The coupling conditions result from the  Gauss~constraints -- which are non--Abelian versions of the Gauss~law for electro--magnetism. In particular if one is interested in  flowing to a phase describing states peaked on a homogeneous curvature one would expect a deformation of these Gau\ss~constraints \cite{DeformedGauss}, and thus a corresponding deformation of the coupling conditions.

Assume we have a model of the form \eqref{TV3}, that is actually triangulation invariant in the same way as \eqref{TV1} is. Such models can appear as exact fixed points of the coarse graining flow  -- that is the models are invariant under the coarse graining procedure even if one does not employ a truncation.\footnote{ More precisely the truncation maps, which are projectors, agree  with the amplitudes of certain building blocks see \cite{TimeEvol}.}  Such a fixed point will be the end point of coarse graining flows which start with some subset of models of the form (\ref{TV3}). This subset of models is referred to as phase which is characterized by the fixed points the models flow to.

To understand the coarse graining flow of a space of models as defined in (\ref{TV3}), it is therefore helpful to know which fixed points one can expect. Here we will discuss one family of such fixed points --  given by the TV amplitudes for $\SU(2)_{\tilde \k}$, where $\tilde k < k$. 

The key point here is that the TV amplitudes for $\SU(2)_{\tilde \k}$ can be described within the $\k$--kinematical structure. To this end one sets the amplitudes ${\cal A}'_\tau$ to zero if any of its arguments is $j>\tilde \k/2$ and defines   
\ba
{\cal A}'_\tau( \{j_e\}) \,=\,\, \tilde G(\{j_e\})\,\, , \q  v'(j_e)\,=\, \tilde v_{j_e}
\ea
otherwise. Here $\tilde G$ and $\tilde v^2_j$ are the $6j$ symbol and the signed quantum dimensions for $\SU(2)_\tk$.  We set the amplitude to vanish if any of the $\tilde \k$--coupling conditions are violated. These conditions do coincide with thus for level $\k$ except for the last requirement in (\ref{Admissibility}), which however is just stronger for level $\tilde \k$ than for level $\k$, as $\tilde \k<\k$.  Thus the amplitudes vanish also if the $\k$--coupling conditions are satisfied and we can therefore express the $\tilde \k$--model within the $\k$--kinematical structure.  We can say that we indeed encounter deformed Gauss~constraints but that this deformation just leads to a stronger version of the original constraints.

In the following we will give a Hamiltonian description of the TV models. More precisely we will define Hilbert spaces and Hamiltonians for which the TV amplitudes  
arise as ground or vacuum states \cite{TimeEvol}.  The important point is that we can use a $\k$ kinematical structure, i.e. a Hilbert space based on $\SU(2)_\k$  and construct a Hamiltonian that has the $\tk$ amplitude as ground state. This allows to consider quantum phase transitions between phases characterized by different values for the level $\k$. Another application is that the $\tk$ vacuum states define an interesting family of states peaked on different values for the homogeneous curvature. Describing homogeneous quantum geometries such states are very suitable for deriving cosmological predictions from quantum gravity.

\section{Hamiltonian description}\label{canonical}

We will now define two different Hilbert spaces, a physical  $\k$ Hilbert space and a kinematical $\k$ Hilbert space. These can be  both  understood to be  Hilbert spaces associated to the TV$_{\k}$ model in the following sense:  The TV$_{\k}$ partition function (with a boundary) can be defined to act as an operator on the kinematical Hilbert space. This operator will be a projector and the image of this projector defines the physical Hilbert space.


\subsection{Physical Hilbert space}\label{physicalHS}

We will define the physical $\k$ Hilbert space via the string net description \cite{LevinWen, KKR, Kir, DG16}.  Given a (for now closed) surface $\Sigma$ this Hilbert space  ${\cal H}_\k$ is spanned by states which are represented by three-valent graphs embedded into the surface $\Sigma$ and whose strands are labeled by $\SU(2)_k$ irreps. The labeling should be such that the coupling conditions (\ref{Admissibility}) are satisfied for the triple of irreps  associated to the three-valent nodes.

Readers familiar with loop quantum gravity will notice, that these states bear some resemblance to the ($q$--deformed) spin network basis states \cite{SmolinLambda}. In loop quantum gravity one would  proceed by defining diffeomorphism and Hamiltonian  constraints and  find solutions to these constraints. The span of these solutions, equipped with an inner product, would then define the physical Hilbert space.

Following the string net description, we will however proceed in a different way. The physical Hilbert space is constructed by imposing equivalence relations on the graph states.  We will see that these equivalence relations do equate states that are related by a time evolution, which in diffeomorphism invariant theories such as gravity is a gauge transformation. 

~\\
Two graph states are equivalent if they can be connected by the following operations:
\begin{itemize}
\item The strands  of a graph can be deformed isotopically within $\Sigma$:
\be\label{StrandDeformation}
\pic{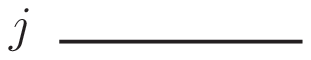}\,=\,\pic{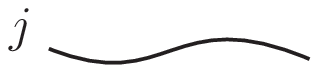}.
\ee
\item Strands with representation $j=0$ can be added or removed:
\be\label{0add}
\pic{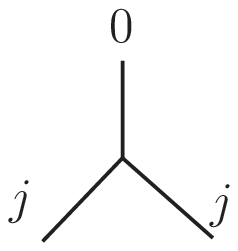}\,=\,\pic{Fig/HLine-j}.
\ee
\item The local connectivity of a graph can be changed by $2-2$  (also referred to as $F$--move) and $3-1$ moves and their inverses:
\ba\label{Fmove}
\pic{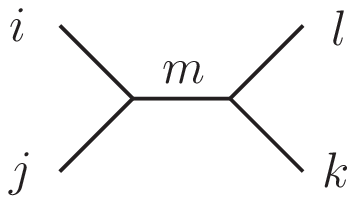}\,&=&\,\,\sum_nF^{ijm}_{kln}\pic{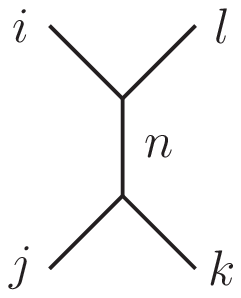} \q ,\\
\label{31move}
\pic{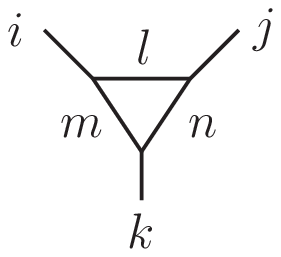}&=&\f{v_nv_l}{v_j}F^{ikj}_{nlm}\pic{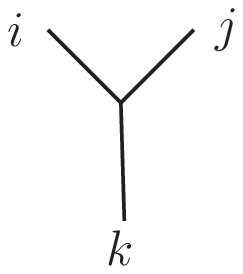} \q .
\ea
\end{itemize}
The $F$--symbol $F^{ijm}_{kln}$ as well as the square roots of the signed quantum dimensions $v_j$ are defined in appendix \ref{SU2Ess}.  The consistency of this set of equivalence relations follows from certain properties of the $F$--symbol listed in appendix \ref{SU2Ess}. Furthermore one can derive from the equations (\ref{Fmove},\ref{31move}) other equivalences which will be very handy for the manipulation of graph states, as we will do below. 

The relations (\ref{Fmove},\ref{31move}) can be understood to arise from an equivalence under time evolution in the following way: As the graphs are three-valent  we can interpret their duals as triangulations. The edges of these triangulations carry the irreps induced from their dual strands. Thus we have now states  given by  superpositions  of two--dimensional labeled triangulations. These can  serve as initial states for a time evolution via the dynamics imposed by  the TV$_\k$ partition function.  In this context the $2-2$, $3-1$ and $1-3$ moves are the basic time evolution moves for a two--dimensional triangulation, resulting from gluing tetrahedra in different ways to the surface \cite{DittrichHoehn2}.

To complete the definition of the physical Hilbert space we need to give an inner product. This can be defined by choosing an independent (with respect to the equivalence relation) and complete set of states and by declaring these states to form an orthonormal basis.  Alternatively, if $\Sigma$ is a (punctured) sphere we can use the trace inner product \cite{KKR,DG16}.

Any triangulation of the sphere can be transformed via the Pachner moves to a triangulation with only two triangles whose edges are pairwise identified. The corresponding dual graph is given by two nodes connected to each other by three strands. The equivalences (\ref{31move}) and (\ref{0add}) allow to reduce this graph even to an empty graph -- corresponding to a totally degenerate triangulation:
\ba\label{ThetaCalc}
\pic{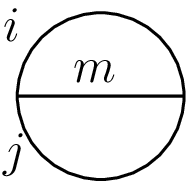}\,\,\,&=&\,\,\, \pic{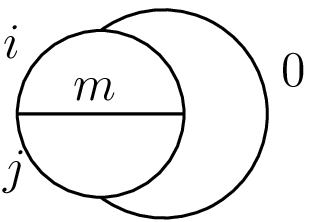}\,\,\,\nn\\&=&\,\,\, F^{ii0}_{mmj} v_m^2 \,\, \, \pic{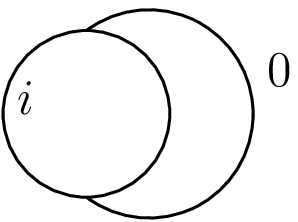} \,\,\,=\,\,\, F^{ii0}_{mmj} v_m^2 \,\,\,\pic{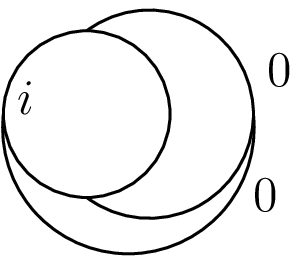} \nn\\
&=&\,\,\,  F^{ii0}_{mmj} v_m^2  F^{000}_{iii} v_i^2 \,\,\,\pic{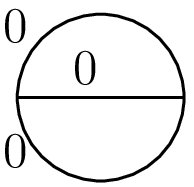} \nn\\
&=&\,\,\, \delta_{mij}\, v_m v_i v_j \q . 
\ea
In the first line we added a zero strand, in the second line we used a 3--1 move and then added again a zero strand, and in the third line we used again the 3--1 move. We remain with a graph which has only strands labelled by $j=0$ representations. According to (\ref{0add}) such a graph is equivalent to the empty graph. We remain with the coefficient of this empty graph  $F^{ii0}_{mmj} v_m^2  F^{000}_{iii} v_i^2$, which due to (\ref{F-normalization}) gives the last line of (\ref{ThetaCalc}). Here the so--called fusion coefficients $\delta_{mij}$ are equal to $1$ if the triple $(m,i,j)$ is admissible, i.e. satisfies the coupling conditions (\ref{Admissibility}) and vanish if this is not the case.

Thus any labelled graph on the sphere (without any punctures) is equivalent to a $\mathbb{C}$--number times the empty graph. This $\mathbb{C}$--number  is called the evaluation of the labelled graph.  Indeed there is only one physical state for the sphere and the physical Hilbert space is thus equivalent to $\mathbb{C}$. As we will discuss in the next section one can introduce punctures on the sphere which will lead to a larger Hilbert space. The reason is that one does allow the equivalences  (\ref{StrandDeformation}--\ref{31move}) only away from the punctures. In particular one is not allowed to move a strand across a puncture (without having a vacuum loop around it) and is also not allowed to use the 3--1 move (\ref{31move}) if a puncture is located in the triangular face appearing in the graph on the left hand side of (\ref{31move}).

We will also employ the following formulas for joining two parallel strands and resolving bubbles:
\ba\label{Joining}
\pic{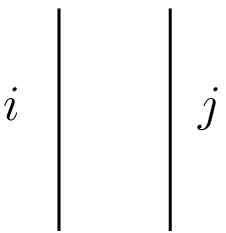}\,\,=\,\,\pic{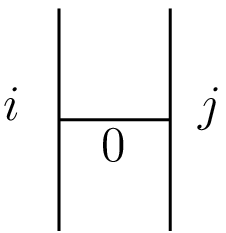}\,\,=\,\, \sum_l F^{ii0}_{jjl}  \,\,\,\pic{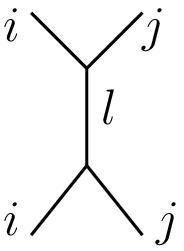} \,\,=\,\,  \sum_l \frac{v_l}{v_i v_j}  \,\,\,\pic{Fig/ParallelLinesC} \q .
\ea
After the 2--2--move  we used a special value for the $F$--symbol, see (\ref{F-normalization}). For resolving the bubble we use the same special value for the $F$--symbol, which this time arises from a 3--1 move:
\ba\label{BubbleMove}
\pic{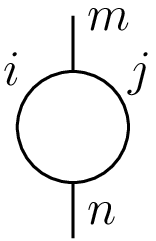}\,\,=\,\,\pic{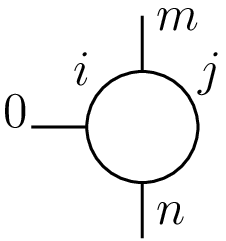}\,\,=\,\, F^{nm0}_{iij} \,v_i^2 \,\,\, \pic{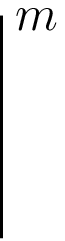}\,\,=\,\, \delta_{mn} \delta_{ijm}  \frac{  v_iv_j}{v_m} \,\,\, \pic{Fig/Line-m}  \q .
\ea

~\\
For the efficient manipulation  of graph states it will be  very useful to introduce two further concepts:

\begin{itemize}
\item We will allow for the crossing of two strands but have to denote which strand is over--crossing and which is under--crossing.  Certain types of double crossings can be resolved by isotopic deformation of the strands, e.g.
\ba
\pic{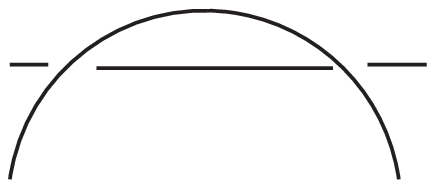}\,\,\,=\,\,\,\pic{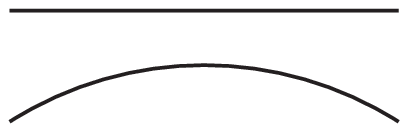} \q .
\ea
In general a crossing is resolved with the $R$--matrix (detailed in appendix \ref{SU2Ess}) by using the following equivalences between graph states:
\be\label{braiding}
\pic{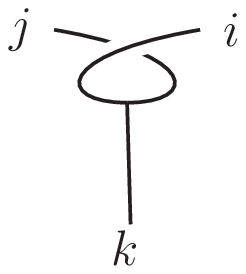}=\,\,R^{ij}_k\pic{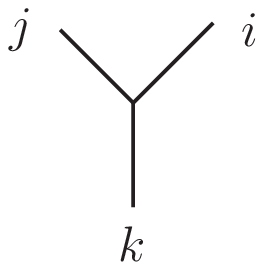},
\q\q
\pic{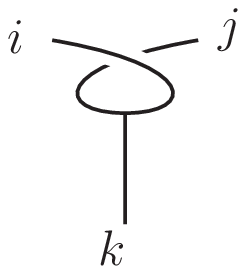}=\,\,\big(R^{ij}_k\big)^*\pic{Fig/Ru2}.
\ee
From these relations one can show that crossings can be resolved as follows
\ba\label{resolution of crossing}
\pic{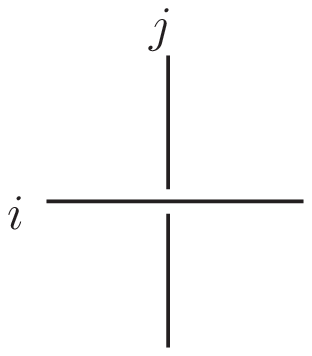}\,&=&\,\,\,\sum_k\f{v_k}{v_iv_j}R^{ij}_k\pic{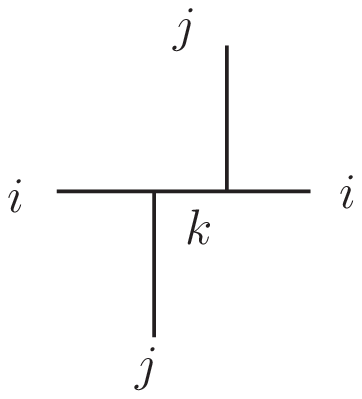},
\,\, \nn\\
\pic{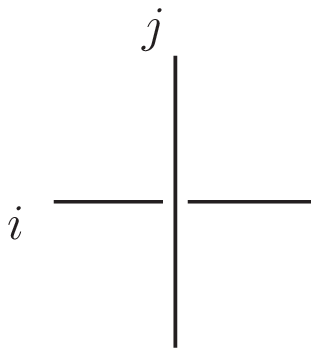}\,&=&\,\,\,\sum_k\f{v_k}{v_iv_j}\big(R^{ij}_k\big)^*\pic{Fig/Uncrossing}.
\ea
This allows to consider a strand encircling another strand
\ba\label{smatrix1}
\pic{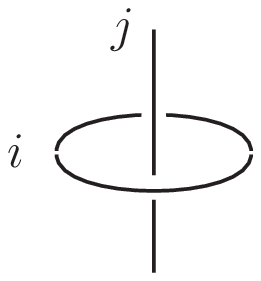}=\,\, \frac{s_{ij}}{s_{0j}}\pic{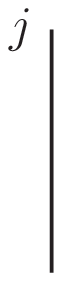} \q .
\ea
Here we have the (rescaled) $S$--matrix appearing which is defined as the evaluation of the following planar graph
\be\label{SM1}
\D \,S_{ij}\,:=\,s_{ij}\,:=\!\pic{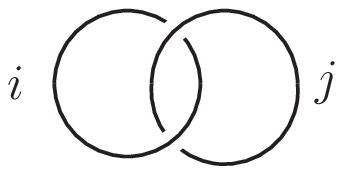}\hs .
\ee
Its explicit value is given in appendix \ref{SU2Ess}. $\D=\sqrt{\sum_j v_j^4}$ is the total quantum dimension.

\item
We furthermore define a short--hand notation for a certain weighted sum over strands labelled by admissible irreps. We will refer to this combination as a vacuum strand and it is defined as 
\be\label{vacuums}
\pic{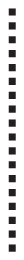} \coloneqq\,\, \,\frac{1}{\D} \sum_k v_k^2 \pic{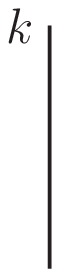}.
\ee
The vacuum strand will actually only appear as vacuum loop. These vacuum loops enjoy the so--called sliding property:
We can slide a strand over a any region (here indicated by a black dot which can stand for an arbitrary complicated graph or a puncture or other topological feature) that is enclosed by a vacuum strand:
\ba\label{sliding}
\pic{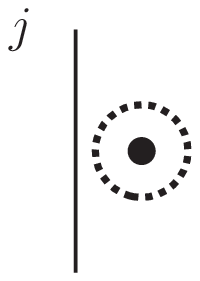}\,\,\,=\,\pic{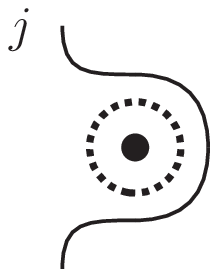}.
\ea
This identity can be proven by using the definition (\ref{vacuums}) for the vacuum strands and the 2--2 move identity (\ref{Fmove}). It shows that the vacuum loop enforces flatness for the (integrated) curvature of the enclosed region in the sense that Wilson lines -- represented by the strands -- can be freely moved over this region. 

\end{itemize}

One can also define operators on the physical Hilbert space, that is operators which are consistent with the equivalence relations  (\ref{StrandDeformation}--\ref{31move}). For a surface without punctures a generating basis of such operators is given by closed ribbon operators.  In the gravitational context these operators describe Dirac observables.  These closed ribbon operators insert an over--crossing and an under--crossing strand  along a closed path into the graph state and are therefore labelled by a pair $(a,b)$ of irreps. The operators can be understood as a product of two parallel Wilson loop operators associated to the two  connections $A_+$ and $A_-$, which we discussed in the introduction.

For surfaces with punctures one can also define open ribbon operators, which describe (exponentiated) electric flux operators, see \cite{DG16} for details. Related observables can also be defined in the covariant framework \cite{BarrettObs}.

\subsection{Kinematical Hilbert space via introduction of punctures} \label{kinHil}

As we have discussed the physical Hilbert space for the sphere is one--dimensional. We can however enlarge the Hilbert space associated to a surface $\Sigma$ by introducing punctures into $\Sigma$, that is by removing disks from $\Sigma$. The equivalence relations (\ref{StrandDeformation}--\ref{31move}) are only allowed away from the punctures. This prevents the reduction of certain graphs, in particular if there are punctures situated inside the faces of a graph. One is thus left with a larger set of independent (with respect to the equivalence relations) states and thus a larger Hilbert space.

Let us note that one usually allows also strands to end at the punctures. To this end one introduces a marked point on the boundary of the puncture which specifies the location where the strand can end. A puncture with such a strand (carrying a non--trivial irrep) represents a torsion excitation in the sense that the Gau\ss~constraints, here in the form of the coupling rules (\ref{Admissibility}) are not satisfied at the one--valent node given by the marked point.  Here we will only consider punctures without torsion, that is strands are not allowed to end at the punctures. (See e.g. \cite{DG16} for the generalization to torsion excitations.) We will however consider in section \ref{general} surfaces with boundary (beside the punctures). As part of the boundary conditions we will allow strands to end at a number of specified marked points. This will define a Hilbert space for surfaces with an `extended' boundary, see \cite{DG16} for the general construction of such Hilbert spaces.

Beside torsion the punctures can also carry curvature excitations. The presence of curvature comes from the fact that we cannot deform strands across the punctures -- if this puncture is not surrounded by a vacuum loop. Inserting a vacuum loop around the puncture will project out the curvature excitation and thus impose the flatness (here understood as homogeneous curvature) constraint for this puncture. 

A two--dimensional piecewise homogeneously curved triangulation can carry curvature at its vertices. Given a triangulation $\Delta$ of $\Sigma$ we  we  adopt all the vertices of $\Delta$ as punctures of $\Sigma$.  We will refer to the resulting string net Hilbert space (for which the equivalence relations  (\ref{StrandDeformation}--\ref{31move}) hold away from the punctures) as {\it kinematical} Hilbert space associated to the triangulation $\Delta$. Note that the Hilbert space does only depend on the (embedding of the) vertices of the triangulation and not on the connectivity of its edges.

We can nevertheless consider the graph $\Gamma_\Delta$ dual to the triangulation $\Delta$ as a canonical choice. The set of all admissible labelings of this graph $\Gamma_\delta$ gives a set of independent states and defines the spin network basis.\footnote{We use this term as this basis is very similar to the spin network basis \cite{RovelliSmolinSNW} used in loop quantum gravity. Note however that the Hilbert space considered here is based on a different vacuum state \cite{DG14a,DG16} as compared to the usual Ashtekar--Lewandowski vacuum \cite{AL}. Due to this change of vacuum one has to use a different version of the flux operators \cite{DG14a}, which appear moreover in exponentiated form. These fluxes are realized via the open ribbon operators, see \cite{DG16}.}

As an example we will have for the tetrahedron the following spin network basis
\ba
{\cal W}^{j_1j_2j_3}_{j_4 j_5 j_6}\,=\,\,\,  \pic{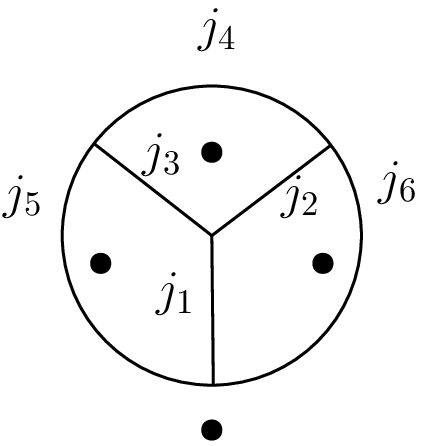} \q ,
\ea
where the punctures are indicated with black dots. 
For a regular triangulation of the torus which has only six-valent vertices we have a graph
\ba
\pic{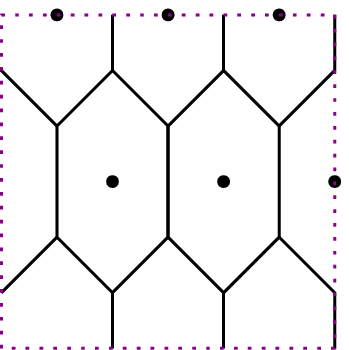}\q, 
\ea
where we have to identify the vertical sides and the horizontal sides of the bounding rectangle to obtain a torus. 

We noted above that the Hilbert space will only depend on the number (and positioning) of the punctures, that is the vertices of the triangulation,  but not on the triangulation itself. We can thus have spin network bases associated to different triangulations as long as these have the same number of punctures. The basis transformation will be given by the $F$ moves (\ref{Fmove}). 

A different type of basis, which is actually dual to the spin network one, is the so--called fusion basis \cite{KKR,DG16}.\footnote{Such a fusion basis can be also defined for undeformed groups \cite{ABC16a}.} See appendix \ref{TTetra} for the definition of the fusion basis for a tetrahedron. The fusion basis is an excitation basis in the sense that a part of the quantum numbers do specify the (curvature or magnetic) excitations of the punctures. These quantum numbers  determine the eigenvalues of closed ribbon operators around the punctures.  This set of quantum numbers is however not complete. Equivalently the set of closed ribbon operators around (single) punctures do not form a maximal set of commuting operators. One has rather to add also ribbons around pairs of punctures and pairs of these pairs and so on. This defines a fusion scheme which can be encoded in a three--valent tree graph whose leaves match the punctures. The eigenvalues for the closed ribbon operators are characterized by two quantum numbers given by a pair of irreps $(a,b)$. In our case we do not allow for torsion excitations, that is for strands ending at the punctures. For this reason we will have $a=b$ for ribbons around single punctures. However one will find in general $a \neq b$ for ribbons around pairs of punctures, showing that torsion can be induced by curvature excitations \cite{EteraCoarse,DG14a}.

We have so far defined (ribbon) operators via the tools of graphical calculus. Having chosen a basis we can also define operators by giving their matrix elements in a basis. We will make use of this fact for the `lifting' of operators from the kinematical $\tk$--Hilbert space to the $\k$--Hilbert space.

\subsubsection{Embedding the kinematical $\tilde \k$--Hilbert space into the kinematical $\k$--Hilbert space}

Given a surface $\Sigma$ fix a triangulation $\Delta$ which determines a dual graph $\Gamma_\Delta$. The spin network basis in the $\k$ Hilbert space consists of all $\k$--admissible labelings  ${\cal W}\{j\}$ of the dual graph $\Gamma_\Delta$.  We can also consider a spin network basis in the $\tk$ Hilbert space --- this basis includes all $\tk$--admissible labelings $\tilde {\cal W}\{j\}$ of the dual graph. A $\tk$--admissible labeling is also $\k$--admissible. We can therefore define an embedding of the kinematical $\tk$ Hilbert space ${\cal H}_\tk$ into the kinematical $\k$ Hilbert space ${\cal H}_\k$ defined by
\ba
{\cal E}_\Delta: \q \,  {\cal H}_\tk \,\, &\rightarrow& \,\, {\cal H}_\k \nn\\
\tilde {\cal W}\{j\} \,&\mapsto& {\cal W}\{j\} \q .
\ea
That is we identify a spin net basis state in ${\cal H}_{\tk}$ labelled with the set of irreps $\{j\}_{{\rm link} \in \Gamma_\Delta}$ with the spin net basis state in ${\cal H}_\k$ which has the same labeling $\{j\}_{{\rm link} \in \Gamma_\Delta}$. We extend this embedding from the spin network basis states to the full Hilbert space ${\cal H}_\tk$ by linearity. 

Note that this embedding does depend on the choice of triangulation, that is in particular on the connectivity of the edges. (In contrast the Hilbert space ${\cal H}_\tk$ and ${\cal H}_\k$ only depend on the number and position of the vertices of the triangulation.)

Having such an embedding we can also map operators ${}_\tk\!\tilde {\bf O}$ on ${\cal H}_{\tk}$ to operators ${}_\k\!\tilde {\bf O}$ on ${\cal H}_k$. The operator 
${}_\k\!\tilde {\bf O}$ is defined to have the same matrix elements with respect to the $\tk$--admissible subset of the spin network basis as the operators ${}_\tk \!\tilde {\bf O}$. Matrix elements of ${}_\k\!\tilde {\bf O}$ involving a $\tk$--non--admissible spin network are defined to be zero. 

This mapping of operators from the ${\cal H}_{\tk}$ Hilbert space to the ${\cal H}_k$ Hilbert space does depend again on the choice of triangulation.

\section{Example: the two-punctured sphere }\label{2punct}

To illustrate the main points we discuss the simplest possible example --- the two--punctured sphere. Note that this two--punctured sphere can be associated to a (very degenerate) triangulation given by one triangle. Two edges of this triangle are glued together -- and can carry any $\SU(2)_\k$ irrep $j$. The third edge is assigned a vanishing length and thus carries the trivial representation. 

Accordingly, an independent basis of states is given by
\ba\label{SQstates}
{\cal W}^j\, :=\, \pic{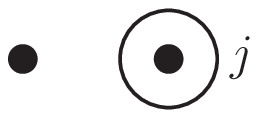}\hs.
\ea
These states form an independent set and can thus be adopted as orthonormal\footnote{The states are also orthonormal with respect to the trace inner product \cite{KKR,DG16}.} basis. (Note that it does not matter around which puncture we draw the $j$--loop as we can slide the loop from one to the other puncture on the sphere.)

As mentioned  we can use the TV$_\k$ partition function to define a projection operator whose (here one--dimensional) image would define the physical state for the sphere. Such projection operators can be even defined locally for a given puncture using so--called tent moves \cite{DittrichHoehn2,DG16}. The resulting projection operator associated to a puncture $p$ is given by: 
\ba\label{Projector}
{\bf B}\,\triangleright   \pic{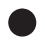}\,\,=\,\,\f{1}{\D}\pic{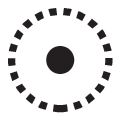}\q ,
\ea
that is, we introduce a (normalized) vacuum loop around the puncture $p$.  We then demand for physical states
\ba
{\bf B} \triangleright  \psi_{\text{phys}} \,=\, \psi_{\text{phys}}
\ea
for all punctures.

The projection operator ${\bf B}$ has a simple action on the states (\ref{SQstates}) for the two--punctured sphere
\ba
{\bf B} \triangleright {\cal W}^j &=& \frac{1}{\D^2} \sum_l v_l^2\pic{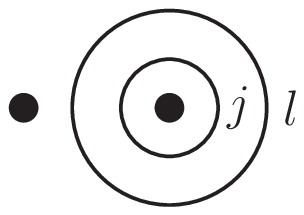}\hs \nn\\  &=& \frac{1}{{\D}^2}  \sum_{ml} v^2_l \delta_{jlm}  \,\, \pic{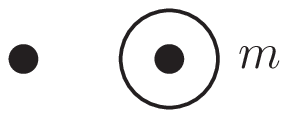}\nn\\
 &=& \frac{1}{{\D}^2}  \sum_{ml} v^2_l \delta_{jlm}  \,\,  {\cal W}^m \q .
\ea
where  we merged two parallel strands into one strand using \eqref{Joining}   and removed a bubble using \eqref{BubbleMove}. 

The one dimensional physical Hilbert space for the two--punctured sphere, which results from applying this projection operator to one of the punctures (as this also imposes the projection for the other puncture) is spanned by the  $\k$--vacuum state
\ba
{\cal O}^0 \,=\, \pic{Fig/Puncture} \pic{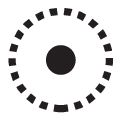}    \,=\,\,\, \frac{1}{\D} \sum_l v_l^2 \, {\cal W}^l \q .
\ea

This state is  one  member of  an alternative basis -- the fusion or Ocneanu basis -- $\{{\cal O}^j\}_{j=0}^{\k/2}$ 
\ba\label{Ostates}
{\cal O}^j \,=\,  \pic{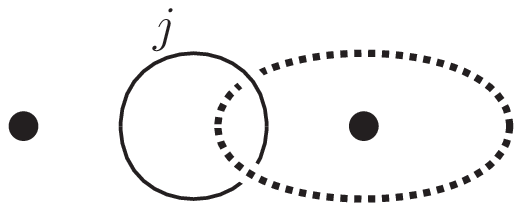}       \,=\,     \sum_{l} S_{jl} \,{\cal W}^l   \q , \q\q {\cal W}^l=\sum_j S_{mj} \,{\cal O}^j \q .
\ea
The matrix $S_{jl}$ is unitary and referred to as $S$--matrix, see appendix \ref{SU2Ess} for its explicit definition.

This basis  diagonalizes the operators ${\bf B}^l$, defined as inserting a Wilson loop in the representation $l$ around the puncture $p$:
\ba
{\bf B}^m \triangleright  {\cal O}^j \,\,&=&\, \, \pic{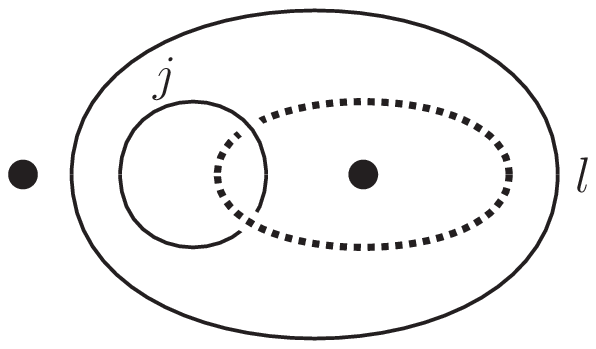}\hs    \q\,=\, \pic{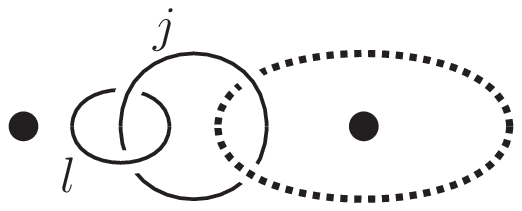}   \nn\\ \,\,  &=& \, \,\,\frac{S_{mj}}{S_{0j}} \,\,  \pic{Fig/Oii00} \,\,\,=\,\,  \frac{S_{mj}}{S_{0j}} {\cal O}^j   \q .
\ea
(To see this use the sliding property (\ref{sliding}) and then the graphical identity (\ref{smatrix1}).)

Let us confirm with an algebraic calculation that the Ocneanu states, given in (\ref{Ostates}) as linear combination of the ${\cal W}$ states,   diagonalize the Wilson loop operators, which in the ${\cal W}$ basis act as
\ba
{\bf B}^j \triangleright {\cal W}^m \,=\, \sum_l \delta_{jml} {\cal W}^l \quad .
\ea 
Thus the ${\cal W}$--basis matrix elements of the Wilson loop operator ${\bf B}^j$ are given by $({\bf B}^j)_{ml}=\delta_{jml}$. The statement then follows from applying the Verlinde formula
\ba\label{Verlinde}
\delta_{jml}&=& \sum_n     S_{mn}  \frac{ S_{jn} }{S_{0n}}   S_{nl} \q ,
\ea
which can be also read as giving the diagonalization of the ${\cal W}$--basis matrix elements of the Wilson loop operator.
(We remind the reader that the $S$--matrix for $\SU(2)_\k$ is real and symmetric as well as unitary.)

Comparing the explicit value of the $S$--matrix in the limit of large $\k$ with the eigenvalues of the $\SU(2)$  Wilson loop operator one sees that one can interpret the label $j$ in ${\cal O}^j$ as (excess) curvature angle $\epsilon\sim\pi(2j+1)/(\k+2)$, see \cite{DG16}.   In the gravitational interpretation such states arise from having a point particle with mass at the puncture. The relation between mass and induced deficit angle is given by $m=\epsilon/8\pi G$  \cite{deSoussa}.  Allowing also for violation of the coupling conditions (that is the Gau\ss~constraints), point particles are characterized by two labels, mass and spin. These labels also arise as representation labels for the irreps of the Drinfel'd Double of $\SU(2)$. (See  \cite{Meusburger} for discussions relating the Drinfel'd Double to point particle excitations in the 3D quantum gravity context.) For $\SU(2)_k$ the Drinfel'd Double is indeed a double, that is given by a tensor product of fusion categories $\SU(2)_\k \otimes (\SU(2)_\k)^{\star}$, where the $\star$ indicates that one uses as $R$ matrix the complex conjugate of the usual $\SU(2)_\k$ $R$--matrix.  Correspondingly, the defect excitations, also referred to as anyons, are labelled by a pair of $\SU(2)_\k$ irreps $(j,j')$.  For a non--spinning particle we have $j=j'$ and the states ${\cal O}^j$ define such excitations.  That is the states ${\cal O}^j$ provide an excitation basis for the Hilbert space of gauge invariant states for the two--punctured sphere.

~\\
We now discuss how to simulate the $\tk$ system on this $\k$ Hilbert space. In the $\tk$ Hilbert space we have also a spin network basis of states
\ba
\tilde {\cal W}^{ j} \,=\, \pic{Fig/Qjj00}\hs.
\ea
where $j\in \{0,\frac{1}{2}, \ldots, \frac{\tk}{2}\}$. The basic idea is to express operators ${}_{\tk}\!\tilde {\bf O}$ on the $\tk$ Hilbert space in this spin network basis and to adopt its matrix elements $\tilde {\bf O}_{ij}$ as the matrix elements of an operator ${}_{\k}\!\tilde {\bf O}$ on the $\k$ Hilbert space.  For the latter operator matrix elements $\tilde {\bf O}_{mn}$ with either $m> \tk/2$ or $n> \tk/2$  are set to zero. (As we are typically working on the $\k$ Hilbert space we  omit in the following the pre-subindex ${}_\k$.)

We thus define the $\tk$ projection operator as
\ba
\tilde {\bf B}  \triangleright {\cal W}^l \,=\, \frac{1}{ \tilde{ \D}^2} \sum_{n,m} \tilde v^2_n \tilde \delta_{lnm} \, \, {\cal W}^m  \q .
\ea
with the understanding that $\tilde \delta_{lnm}=0$ for either $l,n$ or $m> {\tk}/2$. It is defined to vanish for $l+n+m> \tk$. We also define $\tilde v_n=0$ for $n>{\tk}/2$. In general we will set the entries of tilde symbols, such as $\tilde F$ and $\tilde S$, to be vanishing for spin configurations which are not allowed by the $\tk$  coupling rules.

Similarly to the $\tk$ projection operator we can define $\tk$  Wilson loop operators as 
\ba
\tilde {\bf B}^j  \triangleright {\cal W}^l \,=\,  \sum_{m}  \tilde \delta_{jlm} \, \, {\cal W}^m  \q 
\ea
for $j\leq \frac{\tk}{2}$.

Note that $\tilde {\bf B}^j$  -- due to the appearance of $\tilde \delta_{jlm}$ instead of $\delta_{jlm}$ -- is not given by a Wilson loop insertion on this $\k$ Hilbert space, and thus is also not diagonalized by the fusion basis $\{{\cal O}^l\}$.  (Indeed the $\tilde {\bf B}^j$ operators rather measure the curvature with respect to a redefined connection $\tilde A_\pm$ as discussed in the introduction.)  

Instead, we can define the set of states (for $m\leq \frac{\tk}{2}$)
\ba
\tilde {\cal O}^m\,=\, \sum_n \tilde S_{mn}  \, {\cal W}^n  
\ea
 which satisfy
 \ba
 \tilde {\bf B}^j  \triangleright  \tilde {\cal O}^m \,=\, \frac{\tilde S_{jm}}{\tilde S_{0m}} \,  \tilde {\cal O}^m \q .
\ea
As before this follows from the Verlinde formula (\ref{Verlinde}), but now for the $\tk$--fusion rule and $\tilde S$--matrix. 
The states $\{\tilde {\cal O}^m\}_{m=0}^{\tk/2}$, completed with the set   $\{{\cal W}^n\}_{n=\tk+1/2}^{\k}$, form an (orthonormal) basis for the $\k$  Hilbert space. In this sense we can simulate the $\tk$ excitation on the $\k$ Hilbert space.

We can now understand the $\tk$ vacuum and the $\tk$ anyons in terms of the $\k$ anyons and $\k$ vacuum:
\ba\label{ktk1}
\tilde {\cal O}^l \,=\,   \sum_{n,m} \tilde S_{ln}\, S_{nm}        \,\,  {\cal O}^m  \,=:\, \sum_m  M_{lm}  \,\, {\cal O}^m   \q .
\ea
The matrix entry $M_{lm}$ gives the coefficient of the $\k$--anyon $(m,m)$ in the superposition which describes a $\tk$--anyon $(l,l)$.  
E.g. for $\k=3$ and $\tk=2$ we have
\ba\label{M01}
M=
\left(
\begin{array}{cccc}
 0.912 & -0.378 & -0.148 & -0.0613 \\
 0.162 & 0.688 & -0.688 & -0.162 \\
 0.0613 & 0.148 & 0.378 & -0.912 \\
\end{array}
\right) \q ,
\ea
and for $\k=10$ and $\tk=5$ 
\ba
&&M=\nn\\
&&\!\!\!\! \left(
\begin{array}{ccccccccccc}
 0.70 & -0.68 & 0.17 & 0.10 & -0.019 & -0.053 & -0.011 & 0.029 & 0.021 & -0.013 & -0.024
   \\
 0.24 & 0.18 & -0.72 & 0.60 & -0.061 & -0.14 & -0.025 & 0.066 & 0.045 & -0.029 & -0.052
   \\
 0.14 & 0.083 & -0.16 & -0.33 & 0.76 & -0.48 & -0.057 & 0.13 & 0.081 & -0.049 & -0.086 \\
 0.086 & 0.049 & -0.081 & -0.13 & 0.057 & 0.48 & -0.76 & 0.33 & 0.16 & -0.083 & -0.14 \\
 0.052 & 0.029 & -0.045 & -0.066 & 0.025 & 0.14 & 0.061 & -0.60 & 0.72 & -0.18 & -0.24 \\
 0.024 & 0.013 & -0.021 & -0.029 & 0.011 & 0.053 & 0.019 & -0.10 & -0.17 & 0.68 & -0.70
   \\
\end{array}
\right)\!\! .\,\,\,\,\,\,\,\,\,\,\,\,\,\,
\ea
 From this and other examples one can see that in general the $\tk$ vacuum  and the $\tk$ excitations are  superpositions with non--vanishing coefficients for all $\k$  anyons.

The set of sates $\{\tilde {\cal O}^l\}_{l=0}^{\tk/2}$ can be completed to an orthonormal basis in the $\k$ Hilbert space by adding the states $\{{\cal W}^m\}_{m=(\tk+1)/2}^{\k/2}$.   The transformation matrix between the $\k$ fusion basis and this new basis is then provided by the unitary matrix:
 \ba
{\mathbb M}_{mn} =
\begin{cases}
M_{mn} =\sum_{l=0}^{\tk/2} \tilde S_{ml} S_{ln} \q &\text{for}\,\,\, m\leq \tk/2 ,\\
S_{mn}       \q\q  &\text{for}\,\,\,  \tk/2 <m\leq \k/2 \q .
\end{cases}
\ea

The matrix $M$ encodes also to which extend the $\tilde {\bf B}^j$ operators (and therefore the $\k$--projection $\tilde {\bf B}$) fail to be diagonal in the $\k$--fusion basis
\ba
\tilde {\bf B}^j  \triangleright {\cal O}^m \,=\,  \sum_{l,n}     (M^{\rm T})_{m l} \,\, \frac{\tilde S_{jl}}{\tilde S_{l0}} \,\, M_{ln} \, \, {\cal O}^n \quad .
\ea
In particular we have for the $\tilde {\bf B}$ projection operator
\ba\label{OtB1}
\tilde {\bf B} \triangleright {\cal O}^m \,=\,      \sum_{j, l,n}     (M^{\rm T})_{m l} \,\, \frac{\tilde v_j^2}{\tilde{\cal D}^2} \frac{\tilde S_{jl}}{\tilde S_{l0}} \,\, M_{ln} \, \, {\cal O}^n   \,=\,        \sum_{n}  (M^{\rm T})_{m 0} \, \, M_{0n} \, \, {\cal O}^n  \q ,
\ea
whereas the ${\bf B}$ projector is given in the ${\cal O}$--basis as
\ba\label{OB1}
{\bf B} \triangleright {\cal O}^m \,=\, \sum_{n} \delta_{m0} \delta_{0n}  \, \, {\cal O}^n  \q .
\ea
Indeed $\tilde {\bf B}$ projects onto the $\tk$--vacuum state $\tilde {\cal O}^0= M_{0n} {\cal O}^n$.
For  $(\k-\tk)<<\k$  we will have  $M_{0n} \approx \delta_{0n}$ with small corrections, which are decaying with growing $n$.
Thus the set of states ${\cal O}^l$ with small $l$  provides a good truncation for the diagonalization of $\tilde {\bf B}$ or of any combinations of ${\bf B}$ and $\tilde{\bf B}$.  

On the other hand for $\tk=0$ we have $M_{0n}= v^2_n/{\D}$ and ${\bf B}$ projects onto ${\cal W}^0$, which is maximally disordered  in terms of anyon degrees of freedom. In fact, the transformation between the spin network basis and the fusion basis constitutes a duality transform. The reason is that the $\tilde {\bf B}$ projector in the ${\cal W}$--basis for $\tilde k=0$
\ba
\tilde {\bf B} \triangleright  {\cal W}^m \,\, \underset{\tiny{\tk=0}}{=}\,\, \sum_{n} \delta_{m0} \delta_{0n}  \, \, {\cal W}^n 
\ea
has the same matrix elements as the ${\bf B}$ projector in the ${\cal O}$--basis (\ref{OB1}). At the same time the ${\bf B}$ projector in the ${\cal W}$--basis is given as
\ba
{\bf B} \triangleright {\cal W}^m\,\,= \,\, \sum_n S_{m0} S_{0n} \,\, {\cal W}^n
\ea
which is of the same form as the $\tilde {\bf B}$ projector (\ref{OtB1}) for $\tk=0$ in the ${\cal O}$--basis. 

This represents a self--duality similar to the well--known self--duality of the two--dimensional Ising model.

 Thus if we consider a Hamiltonian which interpolates between the two projectors 
\ba
{\bf H}_{\k,\tk=0}(\alpha) \,=\,  -\alpha {\bf B}  \,-\, (1-\alpha) \tilde {\bf B}
\ea
(with $\alpha \in [0,1]$) we will find the same spectrum for ${\bf H}_{\k,\tk=0}(\alpha)$ and ${\bf H}_{\k,\tk=0}(1-\alpha)$.

This property holds however for any Hamiltonian which interpolates in this way between two projectors on one--dimensional subspaces. (This form of the Hamiltonians  will only appear for the case of the two--punctured sphere, in more general cases we will have sums over projectors associated to the different plaquettes and moreover ground state degeneracy for non--trivial topologies .) For such effectively two--dimensional Hamiltonians one can compute the eigenvalues as function of the parameter $\alpha$ and as function of the overlap of the two normalized vectors onto which the Hamiltonian projects for $\alpha=0$ and $\alpha=1$ respectively. In our case this overlap is given by $M_{00}$. The non--vanishing eigenvalues are then given as  
\ba
\lambda_{\pm}(\alpha) \,=\, -\frac{1}{2} \pm \frac{1}{2} \sqrt{ 4 \alpha (\alpha-1) (1- M_{00} (M_{00})^*) +1} \q .
\ea
The minimal gap is reached for $\alpha=1/2$
\ba
\lambda_+(\tfrac{1}{2})-\lambda_-(\tfrac{1}{2}) \,=\, \sqrt{M_{00} (M_{00})^*} \q .
\ea
Thus the gap is proportional to the overlap of the $\tk$ ground state with the $\k$ ground state.  It will be relatively small for $(\k-\tk)$ large and nearly  one for $(\k-\tk)<<\k$. The following table lists $M_{00}$ for different values of $\k$ and $\tk$:

\ba
\left(
\begin{array}{ccccccccccc}
\stackrel{~}{\k} \diagdown  \underset{~}{\tk}&0&1&2&3&4&5&6&7&8&9\\
1& 0.707 & 1 & ~ & ~ & ~ & ~ & ~ & ~ & ~ & ~ \\
 2&0.500 & 0.854 & 1 & ~ & ~ & ~ & ~ & ~ & ~ & ~ \\
 3&0.372 & 0.688 & 0.912 & 1 & ~ & ~ & ~ & ~ & ~ & ~ \\
 4&0.289 & 0.558 & 0.787 & 0.941 & 1 & ~ & ~ & ~ & ~ & ~ \\
 5&0.232 & 0.459 & 0.672 & 0.845 & 0.958 & 1 & ~ & ~ & ~ & ~ \\
 6&0.191 & 0.385 & 0.577 & 0.748 & 0.882 & 0.968 & 1 & ~ & ~ & ~ \\
 7&0.161 & 0.328 & 0.499 & 0.660 & 0.800 & 0.907 & 0.975 & 1 & ~ & ~ \\
 8&0.138 & 0.284 & 0.436 & 0.585 & 0.722 & 0.838 & 0.925 & 0.980 & 1 & ~ \\
 9&0.120 & 0.248 & 0.384 & 0.521 & 0.652 & 0.769 & 0.866 & 0.939 & 0.984 & 1 \\
 10&0.106 & 0.219 & 0.342 & 0.467 & 0.590 & 0.704 & 0.804 & 0.887 & 0.949 & 0.987 \\
\end{array}
\right) \q .\q
\ea

%
%

~\\
{\bf Remark:}
 In the  quantum gravity context one rather works with  Hamiltonian constraints instead of a proper Hamiltonian. These Hamiltonian constraints generate infinitesimal time evolution -- which is however a gauge transformation. The constraints can be also derived from the tent moves \cite{DittrichHoehn2,BonzomThesis}. (See also \cite{BarrettCraneH} for a derivation of the Hamiltonian using the three--dimensional 3--2 Pachner move.)  But instead of summing over the length of the tent pole (which gives the sum defining the vacuum loop) one gauge fixes it lengths -- often to its smallest possible value given by $j=\tfrac{1}{2}$. In general one has a $j$--`ambiguity' for the definition of the Hamiltonian \cite{PerezHamAmb}, but we see here that this $j$--parameter has a straightforward interpretation. The constraints are then given as 
\ba\label{Constraints}
{\mathbf C}^j:= v_j^{-2} {\mathbf B}^j -   {\mathbb I}  
\ea
and are also known as (quantum deformed) flatness constraints.  The condition for physical states is then
\ba\label{HamConsEqu}
{\mathbf C}^j \psi\,=\, 0 \q  \forall \,\,j \q .
\ea
These Hamiltonian constraints can also be defined for the $\tk$--dynamics:
\ba 
\tilde {\bf C}^j \,:= \tilde v_j^{-2} \tilde {\bf B}^j_p -   {\mathbb I}   \q .
\ea
Here we need however to add the $\tk$--admissibility conditions as constraints, that is the $\tk$--deformed Gauss~constraints.

\section{General triangulations} \label{general}

\subsection{Local considerations}

Let us consider a general lattice with three--valent nodes  and the state around one puncture. If we identify the spin network basis with the lattice this is equivalent to considering the state around one plaquette. 

To this end we can consider the Hilbert space of states defined on a one--punctured disk, which represents the state on and around this plaquette.  If we want to represent a plaquette with $N$ links in its boundary we introduce $N$ marked points on the boundary of the disk and allow strands to end on these points. We then consider the family of states:
\ba\label{basisdisk}
{\cal W}^{\{j\}}_{\{l\}}  \,=\,\,\, \pic{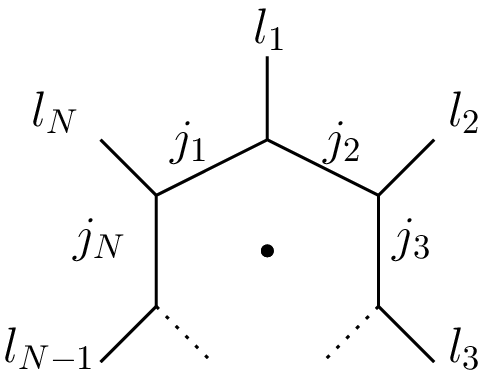}
\ea
Assuming that the set $\{\{j_A\},\{l_B\}\}_{A,B=1}^N$ includes all possible labelings that satisfy the coupling conditions at the nodes  the corresponding  states are independent (with respect to the equivalence relations (\ref{StrandDeformation}--\ref{31move})) and thus we can define an inner product by declaring this set of states to be a orthonormal basis.\footnote{We can also define a trace inner product \cite{KKR,DG16} for this disk Hilbert space -- and the states (\ref{basisdisk}) are also orthonormal with respect to this definition.}
In the following we will only consider the subspace of states with fixed labels $l_1, \cdots, l_N$.

The usual string net projector for such a plaquette is given by
\ba\label{PlaqB1}
{\bf B} \,=\, \frac{1}{{\cal D}^2} \sum_t v_t^2 \, {\bf B}^t
\ea
with
\ba
{\bf B}^t \, \triangleright    \,   {\cal W}^{j_1 \cdots j_N}_{l_1 \cdots l_N}&=& \sum_{m_1, \ldots, m_N}  \left( \prod_{A=1}^N  F^{m_{A+1} l_A m_A}_{j_A t j_{A+1}} \right) \, {\cal W}^{m_1 \cdots m_N}_{l_1 \cdots l_N} \quad ,
\ea
where the index $A=N+1$ is identified with $A=1$.

As in the previous section we can ask for eigenstates of the ${\bf B}^t$ operators. These are given by generalized fusion bases states \cite{DG16}
\ba
{\cal O}^j_{l_1 \cdots l_N, r_1 \cdots r_{N-1}}\,\, \,=\,\,\,\,\pic{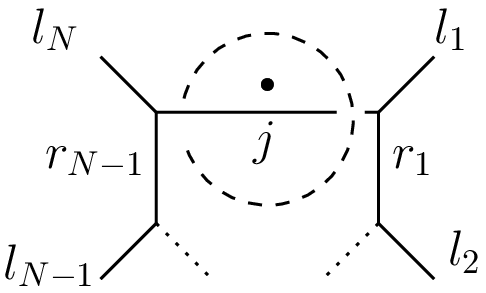}
\ea
which also form an orthonormal basis.  The image of the projector ${\bf B}$ is spanned by the states $ {\cal O}^0_{l_1 \cdots l_N, r_1 \cdots r_{N-1}}$. That is, taking into account the coupling conditions the projector can be written as
\ba\label{PlaqB2}
{\bf B} \,=\, \sum_{ r_1 \cdots r_{N-1}}  \delta_{r_1 l_1} \delta_{r_{N-1} l_N} \left( \prod_{A=1}^{N-2} \delta_{r_A l_{A+1}  r_{A+1}} \right)\; | {\cal O}^0_{l_1 \cdots l_N, r_1 \cdots r_{N-1}} \rangle \langle {\cal O}^0_{l_1 \cdots l_N, r_1 \cdots r_{N-1}}|
\ea
To show that ${\bf B}$ in (\ref{PlaqB1}) and (\ref{PlaqB2}) coincide, one needs to change the bulk of the piece of the triangulation that is glued in the tent move defined by ${\bf B}$ and then use the triangulation invariance of the Turaev--Viro partition function. For $N=3$ this amounts to using the pentagon identity (\ref{Pentagon}), for larger $N$ it defines a generalization thereof. 
 A similar factorization property holds for the projectors onto the anyon states ${\cal O}^j_{l_1 \cdots l_N, r_1 \cdots r_{N-1}} $ with $j>0$.

We see  that for $N\geq 4$ the vacuum state around one plaquette is {\it not} unique, even if we fix the $l$ labels of the links ending in the marked points. We will rather have $(N-3)$ $r$--labels which we can choose freely, as long as the coupling conditions are satisfied.\footnote{In a phase space analysis for N--valent tent moves this corresponds to the fact that one has $(N-3)$ physical degrees of freedom, given by the lengths of the edges adjacent to the $N$--valent vertex minus the three gauge parameters associated to the vertex \cite{DittrichHoehn2}.}

~\\
As in the previous section we can now easily define the projectors and solutions thereof for the $\tilde \k$ dynamics. To this end we put a tilde over all level--dependent entities, e.g. we define 
\ba
\tilde {\bf B}^t \, \triangleright    \,   {\cal W}^{j_1 \cdots j_N}_{l_1 \cdots l_N}&=& \sum_{m_1, \ldots, m_N}  \left( \prod_{A=1}^N  \tilde F^{m_{A+1} l_A m_A}_{j_A t j_{A+1}} \right) \, {\cal W}^{m_1 \cdots m_N}_{l_1 \cdots l_N}\,.
\ea
The image of the corresponding projector 
\ba
 \tilde {\bf B} = \frac{1}{\tilde \D^2} \sum_t \tilde v_t^2 \tilde {\bf B}^t
\ea 
is spanned by states 
\ba\label{PlaqSol}
\tilde  {\cal O}^0_{l_1 \cdots l_N, r_1,r_2  \cdots, r_{N-2}, r_{N-1}} \,=\,   \delta_{r_1 l_1} \delta_{r_{N-1} l_N}\!\! \!\!\!\! \sum_{t,k_1,\ldots,k_{N-1}  }\!\!\!\! \frac{\tilde v_t^2}{\tilde\D} \left( \prod_{A=1}^{N-2}
\tilde F^{k_{A+1}l_{A+1}k_A}_{r_A \,t\, r_{A+2}} \right) \tilde F^{tl_Nk_{N-1}}_{r_{s-1}t0}
 {\cal W}^{tk_1k_2\cdots k_{N-1}}_{l_1\cdots l_N}\nn\\
\ea
keeping in mind that we now have $t,l_A,r_A \leq \tfrac{\tilde \k}{2}$.  

~\\
In a similar manner we can consider the $\k$ and $\tk$ vacuum states around a cluster of adjacent plaquettes. In this case one considers a projector
\ba
{\bf B}_{\rm cluster}\,=\, \prod_p  {\bf B}_p
\ea
where the product is over all plaquettes included in the cluster.
Alternatively one defines the Hamiltonian
\ba
{\bf H}_{\rm cluster} \,=\, - \sum_p {\bf B}_p \q ,
\ea
and looks for its ground states. Here one considers as boundary condition fixed spins $l_A$ associated to the links which end transversally on the boundary around the cluster of plaquettes.  (Similar definitions apply for the corresponding $\tk$--operators.) The solutions can  again be given in terms of generalized fusion bases states.  E.g. for two neighbouring plaquettes we have
\ba
\pic{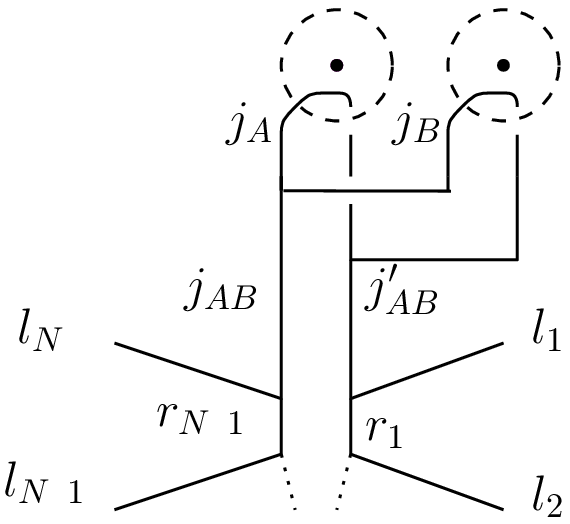} \q ,
\ea
which form a basis for the Hilbert space on the  two--punctured disk with $N$ marked points on the boundary.
Here we encounter quantum numbers that describe the fusion of the anyons $(j_A,j_A)$ on one plaquette  and $(j_B,j_B)$ on the other plaquette  to a fused anyon pair $(j_{AB},j'_{AB})$ where we can now have $j_{AB}\neq j'_{AB}$. These representation labels determine the eigenvalues of ribbon operators going around the two plaquettes. We will discuss such ribbon operators in the next section.

\subsection{Ribbon operators}\label{sec:ribbons}

Here we will discuss how to analyze the $\k$ anyon content  of ta given state, e.g. the $\tk$ vacuum state. This can be done by expanding the global $\tk$ solution into the fusion basis, but this will be very cumbersome for larger lattices. Alternatively we can use (projective) ribbon operators. For one plaquette these will be basically Wilson loops. We can then use the expectation value of these Wilson loops (or of the projective ribbon operators) as an order parameter to distinguish the $\k$ and $\tk$ phase.

Closed ribbon operators can be understood as a generalization of Wilson loop operators. The ribbon operator ${\bf R}_{aa'}$ acts by inserting an over-crossing $a$ strand and a parallel under-crossing $a'$ strand along a closed loop, e.g. 
\ba
  {\bf R}_{aa'} \triangleright    \pic{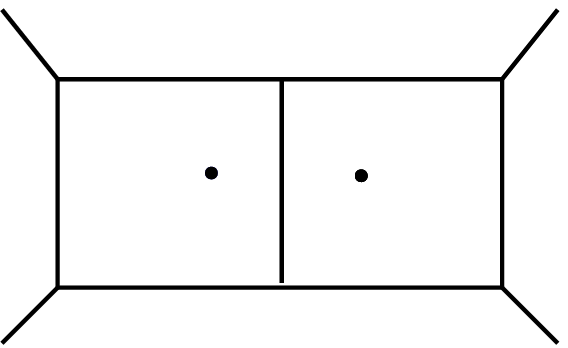}\,\, \,\,=\,\,\,\,\pic{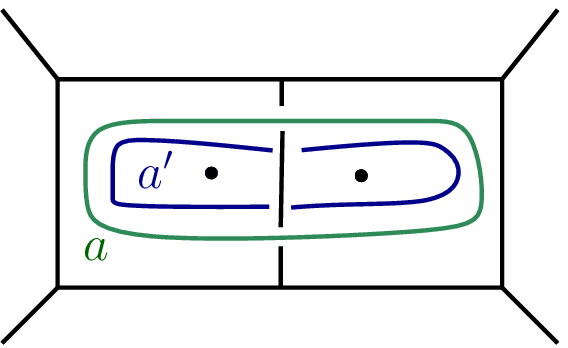}
\ea
  The operator is invariant under isotopic deformation of the underlying loop, i.e. we need to only know how the loop winds around the various punctures. 

The ribbon operators are diagonalized by an appropriately chosen fusion basis and thus allow us to understand a given state as a linear combination of anyon excitations. Projective ribbon operators project onto the subset of the fusion basis with the appropriate anyon labels. They are defined as
\ba\label{ProR}
{\bf P}_{aa'}&=& \frac{1}{{\cal D}^2} v_a^2 v_{a'}^2 \sum_{bb'} S_{ab} S_{a'b'} \, {\bf R}_{bb'}
\ea
and are indeed projectors, i.e. ${\bf P}_{aa'} {\bf P}_{bb'} \,=\, \delta_{ab}\delta_{a'b'}  {\bf P}_{aa'}$. 

If we can deform the loop underlying a projective ribbon ${\bf P}_{ab}$ such that it does not cross any strand of a state, the projective ribbon does reduce to the insertion of one Wilson loop
\ba
{\bf P}_{aa'} \triangleright\!\!\! \pic{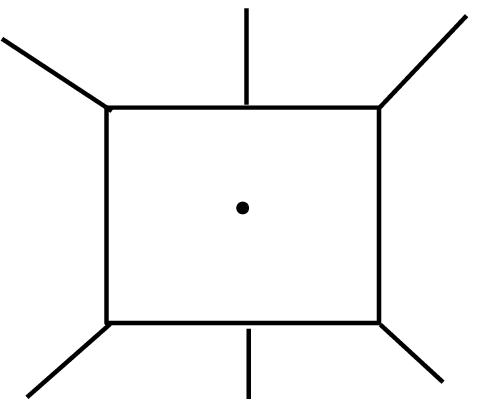}\! \!\!=\,   \frac{1}{{\D}^2} v_a^2 v_{a'}^2 \sum_{bb'} S_{ab} S_{a'b'}  \!\! \pic{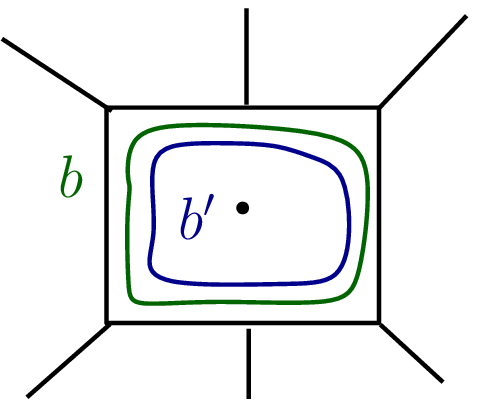} \!\!\! =\,\, \delta_{a a'}  \frac{v_a^2}{\D} \sum_c S_{ac} \!\! \!\!\! \!\!\!\pic{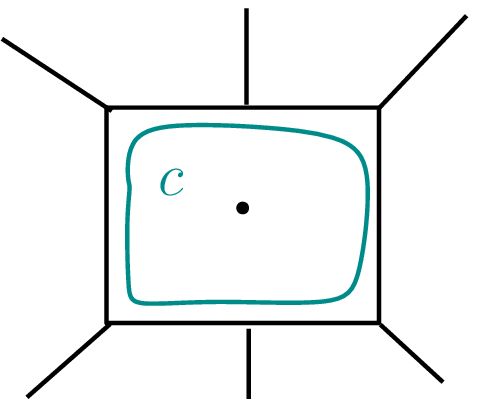}
\ea

We have thus for the projective ribbon around one puncture --- or around one plaquette
\ba
{\bf P}_{aa'} \triangleright  {\cal W}^{j_1 \cdots j_N}_{l_1 \cdots l_N} \,=\, \delta_{a a'}  \frac{v^2_a}{{\D}}  \sum_C S_{ac} \sum_{m_A}  \left( \prod_{A=1}^N  F^{m_{A+1} l_A m_A}_{j_A c j_{A+1}} \right) \,\,{\cal W}^{m_1 \cdots m_N}_{l_1 \cdots l_N}
\ea
In the following table  we give for $k=3$ the expectation values for the operator ${\bf P}_{aa}$ around one puncture for the $(\tk=2)$--vacuum state on a tetrahedral lattice. We thus have a plaquette with three links.
 These expectation values give the probabilities to find an anyon excitation $(a,a)$ at a given puncture:
  \ba
\begin{array}{|c|c|c|c|}\hline
a=0&a=\tfrac{1}{2}&a=1&a=\tfrac{3}{2}\\\hline
0.599&0.383&0.017&0.001\\\hline
\end{array}
 \ea
 Note that these differ from the probabilities for the two--punctured sphere, which are given by the square of the entries in the first row of the matrix (\ref{M01}).  

Next we consider the projective ribbon operators around more  punctures. Using the spin network basis the ribbon will cross a number of strands -- and thus we cannot exclude that (fused) anyons $(a,a')$ with $a\neq a'$ will appear. 

The crossing of a ribbon with a strand can be computed using the definitions \eqref{resolution of crossing} for the $R$--matrix and by reducing any closed face without a puncture  using the equivalences in section \ref{physicalHS}.  One obtains 
\ba\label{ProjRAc}
&&{\bf R}_{aa'}  \triangleright   \pic{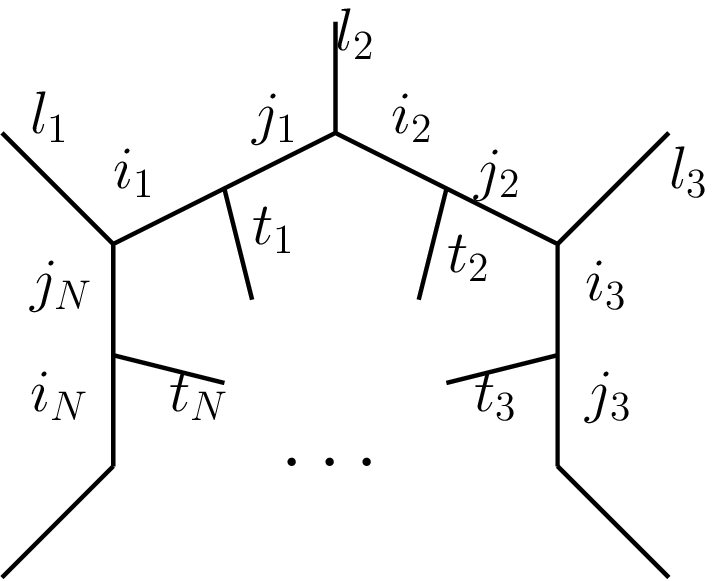}  \nn\\
~\nn\\
&=&\!\!\!\!\!\!
\sum_{r_A,q_A,m_A,n_A}  \!\!\!
\frac{ v_{i_1} \cdots v_{i_N}}{v_{j_1} \cdots v_{j_N}}  v_{t_1} \cdots v_{t_N} \,\,\,
\Omega^{r_1r_2}_{q_1t_1,aa'} \Omega^{r_2r_3}_{q_2t_2,aa'} \cdots  \Omega^{r_Nr_1}_{q_Nt_N,aa'} \,\,\, 
F^{n_Nl_1m_1}_{i_1r_Nj_N}F^{n_1l_2m_2}_{i_2 r_1 j_1} \cdots F^{n_{N-1} l_N m_N}_{i_N r_{N-1}  j_{N-1}}  \nn\\
&&\q
F^{n_1i_1q_1}_{t_1r_2j_1} F^{n_2i_2q_2}_{t_2 r_3 j_2} \cdots F^{n_Ni_Nq_N}_{t_Nr_1j_N}\,\,\, 
  F^{m_1 t_1 n_1}_{q_1i_1r_1}  F^{m_2 t_2 n_2}_{q_2 i_2r_2} \cdots F^{m_N t_N n_N}_{q_N i_Nr_N} \,\,\,
 \pic{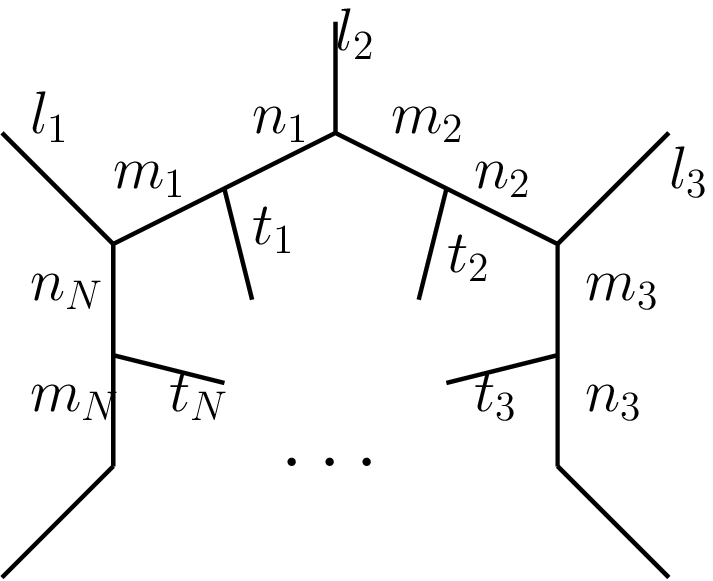}  \nn\\
\ea
where 
\ba
\Omega^{rp}_{ql,ij}&=&\sum_{mn}\f{v_mv_n}{v_rv_l^2}R^{il}_mR^{lj}_nF^{nmr}_{ijl}F^{qlp}_{ijm}F^{rlq}_{jmn}\,\,=\,\, 
\f{1}{v_iv_jv_rv_l^2}\pic{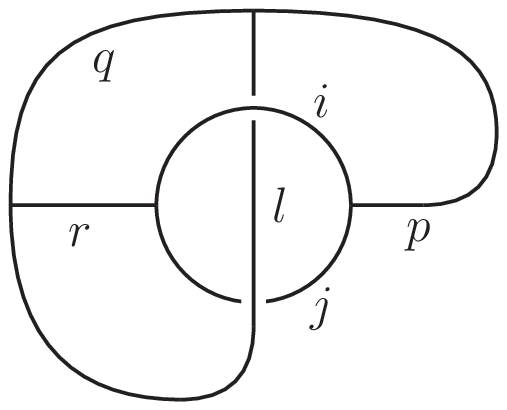}
\ea
is the so--called half braiding tensor. 
(The projective ribbon is then defined by (\ref{ProR}).)  Equation (\ref{ProjRAc}) covers the most general case of a ribbon operator around a number of plaquettes. Special cases can be obtained by setting some labels to the trivial representation, e.g.  $t_B=0$ for some index $B\in {1,\ldots,N}$, in which case one has $i_B=j_B$ and the sum also restricts to $m_B=n_B$.

The following table shows the expectation values for a projective ribbon ${\bf P}_{aa'}$ around two neighbouring plaquettes in a tetrahedral lattice for $(\k=3,\tk=2)$. 
\ba
\begin{array}{|c|c|c|c|c|}\hline
a\diagdown a' & 0 &\tfrac{1}{2}&1&\tfrac{3}{2} \\\hline
0&0.524 & 0 & 0.0365 & 0 \\
\tfrac{1}{2}& 0 & 0.378 & 0 & 0.00261 \\
1& 0.0365 & 0 & 0.0189 & 0 \\
\tfrac{3}{2} & 0 & 0.00261 & 0 & 0.000693 \\\hline
\end{array}
\ea
As one can see, fused anyons $(a,a')$ with $a\neq a'$ indeed appear.

 \eqref{ProjRAc} shows that the action of the ribbon operator will  only depend on the state in a tubular neighbourhood around the ribbon.  
Now the $\tilde \k$ vacuum states on different lattices   can be also related by Pachner moves, similarly to the usual string net equivalence relations in section \ref{physicalHS}. It follows that one will find the same expectation values for ribbons which go around a region that has been changed by such Pachner moves. E.g.
\ba
\pic{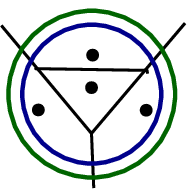}\,\,\sim\,\,  \pic{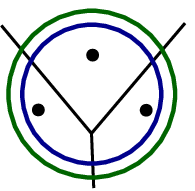} \q, \q\q  \pic{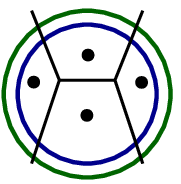}\,\,\sim\,\,  \pic{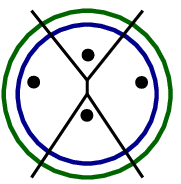}
\ea
Here the $\sim$ sign stands for having the same expectation value for the shown $\k$ ribbon operators in the $\tilde \k$ vacua based on the two different graphs shown on the left and right hand side of the equations respectively.  Thus the $\tk$ vacuum states as expressed in the $\k$ Hilbert space enjoy a certain notion of triangulation independence.

\subsection{Global solutions}

For a general lattice dual to a triangulation we can define the $\tk$--Hamiltonian
\ba
\tilde {\bf H} \,=\,  - \sum_{p} \tilde {\bf B}_p \,\, -c_{\rm link} \sum_{l} \tilde {\bf A}_l  - c_{\rm node} \sum_n \tilde  {\bf A}_n \quad 
\ea
with $c_{\rm link}>0$ and $c_{\rm node}>0$. 
Here we have also introduced operators $\tilde {\bf A}_l$ which act on the links and are equal to the identity if the link carries a spin $j\leq \tk/2$ and give vanishing values otherwise. Likewise $\tilde {\bf A}_n$ are operators on the nodes which project onto states satisfying the coupling condition $j_1+j_2+j_3 \leq \tk$ for this node. If we are only interested in the ground state we can set $c_{\rm link} $ and $c_{\rm node}$ to zero as already the $\tilde {\bf B}_p$ are annihilating states which do not satisfy the $\tilde \k$ coupling condition. Introducing a non--vanishing $c_{\rm node}$ does however change the excitation spectrum by making the violations of the coupling conditions more expensive.

The Hamiltonian has a unique ground state for planar lattices, but has degenerate ground states for non-trivial topology. E.g. for the torus we will have $(\tk+1)^2$ ground states.

The ground states can be constructed with a similar strategy as in section \ref{2punct}. 
That is one expresses first the $\k$ vacua into the spin network basis using the lattice under consideration. One then obtains the $\tk$ vacua by putting a tilde over every term that appears in this expansion. 

E.g. for a tetrahedral lattice we obtain
\ba
\psi_0\,\,=\,\,\,\,\pic{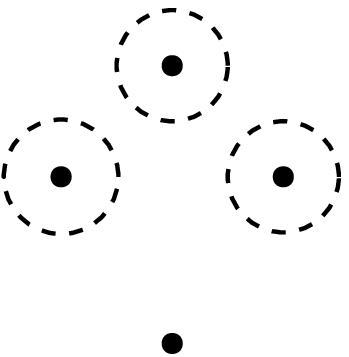}\,\, \,\,=\,\, \frac{1}{{\D}^3}  v_{j_1}v_{j_2} v_{j_4}v_{j_5}    F^{j_1j_2j_3}_{j_4j_5j_6} \,\,\,\,\, \pic{Fig/6jA} 
\ea
and thus for the $\tk$--vacuum
\ba
\tilde \psi_0 \,=\, \frac{1}{\tilde{\D}^3}  \tilde v_{j_1} \tilde v_{j_2} \tilde v_{j_4} \tilde v_{j_5}  \tilde   F^{j_1j_2j_3}_{j_4j_5j_6} \,\,\,\,\, \pic{Fig/6jA}\,\,\,\, .
\ea
For a more general planar  lattice the expansion coefficients coincide with the Turaev--Viro amplitude for a spherical building block with boundary data defined by the spin network state, see section \ref{covariant}. 

Let us also discuss the case of a torus. Allowing for one puncture on the torus we choose a spin network basis
\ba\label{SNWbasisTorus}
{\cal W}^{pqk}&=&\,\,\,\pic{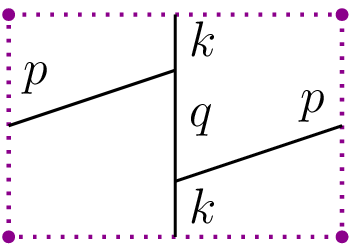} \q .
\ea
Here we have to identity the horizontal and vertical sides of the (dotted) rectangle  to obtain a torus. This torus has one puncture which in the rectangle representation appears in the corners.  Note that this puncture is situated in a  six--valent plaquette.   The set of spin network states defined in (\ref{SNWbasisTorus}) can be also used for the torus without a puncture. The set is, however, in this case  over-complete -- one has one closed plaquette without a puncture inside. Inserting a (normalized) vacuum loop inside  this plaquette will therefore not change the states. But by reducing this vacuum loop in the same way as for the plaquette Hamiltonian one obtains an combination of spin network states which describe a state equivalent to the original one. 

For the torus the vacua are degenerate -- there are $(\k+1)^2$ ground states. The $\k$ vacua for the torus without punctures are given by (see \cite{KKR,BD17})
\ba
\psi^{(0)}_{ab}  \,=\, \pic{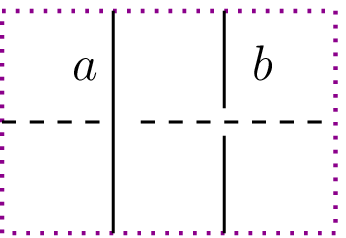} \q .
\ea
We thus have for the one-punctured torus
\ba\label{Torus1Plaqu}
\psi^{(1)}_{ab} \,&=&\,\, \pic{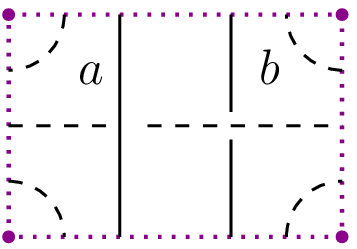}\,\,=\,\, \D\,\, \pic{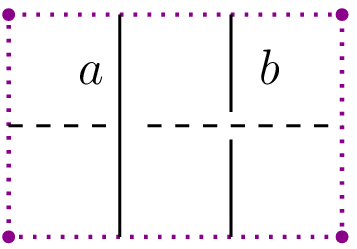}\,\,\,\,=\,\,\,  \D \sum_k \frac{v_k}{v_av_b}\,\,\,  \pic{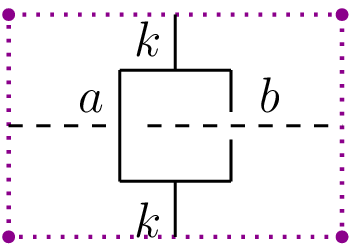} \nn\\
&=& \sum_{k,p,q} v_p^2 \Omega^{kk}_{qp,ab} \,\,\, \pic{Fig/TorusSNW1} \q .
\ea
Here we used in the first equation of (\ref{Torus1Plaqu}) the sliding property: the vacuum loop around the puncture is slid across the `torus hole' using the remaining vacuum loop. The result is that one has two vacuum loops in parallel, which reduce to one vacuum loop and a factor of $\D$.  The resulting state can then be rewritten into a linear combination of fusion basis states\footnote{In this case we allow also for strands to end at the punctures.} for the two--punctured sphere, which we obtain by identifying the two vertical sides of the rectangle. These fusion basis can be expanded into spin network states on the same  two--punctured sphere and in this correspondence we have again the half--braiding tensor appearing, see e.g. \cite{DG16}. 

Similarly we have for the two--punctured torus
\ba\label{Torus2Plaqu}
\psi^{(2)}_{ab} \,&=&\,\,\pic{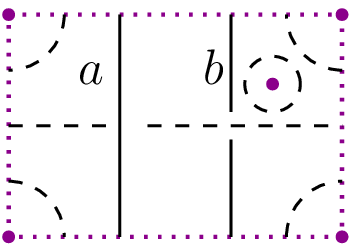}\,\,\,=\,\,\, \sum_{k,p,q} v_p^2 \Omega^{kk}_{qp,ab} \,\,\,\pic{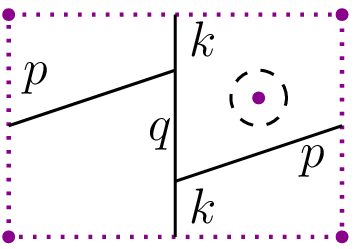} \nn\\  \,\,\,&=&\,\,\, 
\frac{1}{\D}\sum_{k,p,q, j_2,j_3}   \Omega^{kk}_{qp,ab} \frac{v_{j_2} v_{j_1}}{v_k}  F^{j_1 p j_5}_{q j_2 k} F^{j_5 k j_4}_{pj_2 q} F^{j_2 j_4 p}_{j_5 j_1 j_3}
\pic{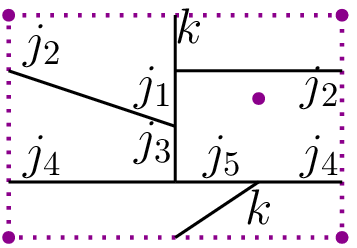} \q .
\ea
Here we expanded the vacuum loop around the additional puncture into $j_2$--loops and merged part of these loops with the $k,p$ and $q$ strands.  The resulting configuration allows for two 3-1 moves which leads to two of the $F$--symbols. A final 2-2 move leads to a spin network state for which both punctures sit in 6--valent plaquettes.

To obtain the solution for a torus with more punctures one proceeds similarly as in going from the one--punctured torus to the two--punctured torus. This can be understood as a refinement operation which can be interpreted as gluing tetrahedra -- coming with the corresponding $F$--symbols and $v$--factors -- to the triangulation of the  torus.

To obtain the $\tk$--vacuum states  we replace all $\k$ dependent expressions in (\ref{Torus1Plaqu},\ref{Torus2Plaqu}) with the corresponding $\tk$--expressions.

For small lattices it is also feasible to expand the $\tk$--vacua into the $\k$--fusion basis states, see  appendix \ref{TTetra} for such an expansion for the tetrahedron. The expansion coefficients encode the anyon content at all the fundamental plaquettes and the fused plaquettes.

\section{Discussion}\label{discussion}

A key difficulty in extracting large scale dynamics from quantum gravity models such as loop quantum gravity is to construct states which approximate well smooth geometries. In this work we defined a family of states that are peaked on homogeneous curvature and analyzed the excitation content of these states.
 The states appear as solutions of anomaly--free  first class constraints. These constraints describe the dynamics of three--dimensional gravity with a cosmological constant. 
 
The same strategy can be applied to the $(3+1)$--dimensional theory. In this case the states are vacua of the Crane--Yetter model for $\SU(2)_\tk$ and one can also define a set of anomaly free first class constraints which describe these solutions.  The states are peaked on homogeneous geometries. These geometries are however generalized in the same way as the  standard loop quantum gravity geometries are, if understood in terms of piecewise flat simplicial building blocks  \cite{DittrichRyan}. 

The transition from the $\k$ vacuum to a $\tk$ vacuum can be seen as a condensation of curvature excitations. A related concept led to the definition of the `new' vacua and associated new Hilbert space constructions for loop quantum gravity peaked on flat connections \cite{DG14a} and homogeneously curved geometries \cite{DG16,BD17}. Fixing a triangulation we showed that different vacua can be also expressed in one and the same Hilbert space associated to this triangulation.  The states do however  enjoy a certain notion of  triangulation independence, as explained at the end of section \ref{sec:ribbons}.

We have already discussed in the introduction a number of applications of this work.

Let us comment more on the possible nature of the phase transitions between $\k$ and $\tk$ vacua. Continuous transition can lead to conformal field theories and are therefore of particular interest.  There does not seem to be much known in general about these transitions. Here we proposed an order parameter, namely the expectation value of $\k$ curvature operators, which vanish in the $\k$ vacua and acquire non--vanishing values in the $\tk$ vacua. 

We can also consider some special cases. As mentioned in the introduction, transitions from $\tk=0$ to $\k>0$ would correspond to the transition from strong to weak coupling of a $q$--deformed lattice gauge theory. For $\k=1$ the system is equivalent to the gauge Ising model (which is dual to the Ising model) and there the transition from weak to strong coupling is of second order. Two non-Abelian examples that have been considered with different techniques \cite{Schulz} are the golden string net\footnote{For the golden string net model only integer spin representations $j=0$ and $j=1$ are allowed and the admissible triples are $(0,0,0)$, $(0,1,1)$ and permutations as well as $(1,1,1)$.}   which is based on $\SO(3)_{3}$  and the $\SU(2)_2$ string net model.   There are strong indications \cite{Schulz} that for  transition from the $\k=3$ phase tin the golden string net to the $\tk=0$ phase and the transition from the $\k=0$ phase of the $\SU(2)_2$ string net  to $\tk=0$ are both of second order. 

The (numerical) investigations of these phase transitions will get very involved with growing $\k$. For small differences $(\k-\tk)$ it might be however profitable to transform to the fusion basis and apply a truncation in this basis. This case might be also treatable with analytical tools by using an expansion of $\tilde q$ around $q$, which is dicussed in \cite{Etera17} for the case $q=1$.

\begin{center}
\textbf{Acknowledgements}
\end{center}
I would like to thank Cl\'ement Delcamp, Marc Geiller, Aldo Riello and Sebastian Steinhaus for discussions. This work is supported by Perimeter Institute for Theoretical Physics. Research at Perimeter Institute is supported by the Government of Canada through Industry Canada and by the Province of Ontario through the Ministry of Research and Innovation.

\begin{appendix}
\section{Essentials on $\SU(2)_\k$}\label{SU2Ess}

\noindent 
Here we will give  some basic facts about  the fusion category ${\rm SU}(2)_{\rm k}$. More background can be found in \cite{Qbackground}.
 ${\rm SU}(2)_{\rm k}$ can also be understood as a quantum deformation of $\SU(2)$. The deformation parameter is given by a root of unity
 \ba
 q=e^{2\pi\i/({\rm k}+2)} \q .
 \ea
 where ${\rm k}$ is a positive integer.
With this deformation parameter we define the quantum numbers
\be\label{qNumber}
[n]\,:=\,\frac{q^{n/2}-q^{-n/2}}{q^{1/2}-q^{-1/2}}=\frac{\sin\left(\tfrac{\pi}{{\rm k}+2}\,n\right)}{\sin\left(\tfrac{\pi}{{\rm k}+2}\right)},
\q
\forall\,n\in\mathbb{N}-\{0\},
\ee
with $[0]=1$. 

The objects of the fusion category ${\rm SU}(2)_{\rm k}$ are given by the admissible irreducible and unitary representations of the corresponding quantum deformed group. These admissible irreps are labeled by `spins' $j\in\{0,1/2,1,\dots,{\rm k}/2\}$.

The quantum dimensions are given by $d_j=[2j+1]$. Admissible representations, that is irreps $j\in\{0,1/2,1,\dots,{\rm k}/2\}$ have positive, non--vanishing quantum dimensions.  It is convenient to introduce the signed quantum dimensions
\ba
v_j^2\,:=\,(-1)^{2j} d_j
\ea
and their square roots $v_j$ (fixing once and for all one root).  The total quantum dimension is defined as 
\ba\label{qdimension}
\D&\,:=\,&\sqrt{\sum_jv_j^4}\,\,=\,\,\sqrt{\f{{\rm k}+2}{2}}  \frac{1}{\sin\left(\f{\pi}{{\rm k}+2}\right)} \q .
\ea

We can now proceed to the recoupling theory for ${\rm SU}(2)_{\rm k}$.  As in the group case we can tensor representations (although with a deformed co--product). In the terms of the fusion category this defines a `fusion' product. 
Admissible triples are triples $(i,j,l)$ of irreps that include the trivial representation in their tensor product. Such triples $(i,j,l)$ are defined by the  conditions:
\be\label{Admissibility2}
i\leq j+l,
\q
j\leq i+l,
\q
l\leq i+j,
\q
i+j+l\in\mathbb{N},
\q
i+j+l\leq{\rm k}.
\ee
The fusion symbol  $\delta_{ijl}$ is equal to one if $(i,j,l)$ is an admissible triple and vanishes otherwise. 

The $F$--symbols transform between different bracketings for the tensor product.  To define the $F$--symbols we introduce first  for any admissible triple $(i,j,k)$ the quantity
\be
\Delta(i,j,k)\,:=\,\delta_{ijk}\sqrt{\f{[i+j-k]![i-j+k]![-i+j+k]!}{[i+j+k+1]!}},
\ee
where $[n]!\coloneqq[n][n-1]\dots[2][1]$.

 The (Racah--Wigner) quantum $\{6j\}$ symbol is then given by the formula
\be
\left\{
\begin{array}{ccc}
i&j&m\\
k&l&n
\end{array}\right\}
:=&\:\Delta(i,j,m)\Delta(i,l,n)\Delta(k,j,n)\Delta(k,l,m)\sum_z(-1)^z[z+1]!\nonumber\\
&\times\f{\Big([i+j+k+l-z]![i+k+m+n-z]![j+l+m+n-z]!\Big)^{-1}}{[z-i-j-m]![z-i-l-n]![z-k-j-n]![z-k-l-m]!},\q
\ee
where the sum runs over
\be
\max(i\!+\!j\!+\!m,i\!+\!l\!+\!n,k\!+\!j\!+\!n,k\!+\!l\!+\!m)\leq z\leq\min(i\!+\!j\!+\!k\!+\!l,i\!+\!k\!+\!m\!+\!n,j\!+\!l\!+\!m\!+\!n).
\ee
The $F$--symbols can then be defined as
\be\label{FDefinition}
F^{ijm}_{kln}
\,:=\,
(-1)^{i+j+k+l} \,\sqrt{[2m+1][2n+1]}\,
\left\{
\begin{array}{ccc}
i&j&m\\
k&l&n
\end{array}\right\}.
\ee
We will also use the $G$--symbol
\ba
G^{ijm}_{kln}
\,:=\, \frac{1}{v_m v_n} F^{ijm}_{kln}\,=\,   (-1)^{i+j+k+l+m+n}   \left\{
\begin{array}{ccc}
i&j&m\\
k&l&n
\end{array}\right\}.
\ea

The $F$ symbol satisfies a number of consistency conditions and properties
\begin{subequations}
\ba
\text{Physicality:}\q&F^{ijm}_{kln}\,=\,F^{ijm}_{kln}\delta_{ijm}\delta_{iln}\delta_{kjn}\delta_{klm},\\
\text{Tetrahedral symmetry:}\q&F^{ijm}_{kln}\,=\,F^{jim}_{lkn}\,=\,F^{lkm}_{jin}\,=\,F^{imj}_{knl}\f{v_mv_n}{v_jv_l},\\
\text{Orthogonality:}\q&\sum_nF^{ijm}_{kln}F^{ijp}_{kln}\,=\,\delta_{mp}\delta_{ijm}\delta_{klm},\label{F-orthogonality}\\
\text{Reality:}\q&\big(F^{ijm}_{kln}\big)^*\,=\,F^{ijm}_{kln},\\
\text{Normalization:}\q&F^{ii0}_{jjk}=\f{v_k}{v_iv_j}\delta_{ijk},\label{F-normalization}\\
\text{Pentagon identity:}\q&\sum_nF^{ijm}_{kln}F^{pql}_{nir}F^{rqn}_{kjs}\,=\,F^{ijm}_{spr}F^{pql}_{kms}. \label{Pentagon}
\ea
\end{subequations}

We furthermore need the $R$--matrix which for  ${\rm SU}(2)_{\rm k}$  is given by
\ba
R^{ij}_k&=&(-1)^{k-i-j}\,\left(q^{k(k+1)-i(i+1)-j(j+1)}\right)^{1/2}  \q .
\ea
The $R$--matrix is subject to the hexagon identity:
\ba\label{hexagon identity}
\text{Hexagon identity:}\q&R^{ki}_mF^{kim}_{ljp}R^{kj}_p \,=\, \sum_{n}  F^{ikm}_{ljn}R^{kn}_lF^{jin}_{lkp}\,\,\, .
\ea

The $S$--matrix  is defined as evaluation of the Hopf knot, see (\ref{SM1}). Explicitly we have
\ba\label{s matrix in terms of R}
S_{ij}\,=\, \frac{1}{\D} s_{ij}\,=\,\frac{1}{\D}\sum_l v_l^2R^{ij}_lR^{ji}_l\,=\,\frac{(-1)^{2(i+j)}}{\D}[(2i+1)(2j+1)].
\ea

The $S$-matrix for ${\rm SU}(2)_{\rm k}$ is invertible and unitary, making $\SU(2)_{\rm k}$ into a modular fusion category. Note that the $S$--matrix is also real and symmetric:
\be\label{s matrix identities}
S_{ij}=S_{ji},
\q\q
\sum_lS_{il}S_{lj}=\delta_{ij}\q .
\ee

\section{Diagonalization of a sum of two projectors onto one--dimensional subspaces}\label{diag2P}

Consider a Hamiltonian operator 
\ba
{\bf H} &=& -\alpha |v\rangle \langle v| \,-\, (1-\alpha)  |w\rangle \langle w| 
\ea
which is built from two projectors onto the normalized vectors $|v\rangle$ and $w\rangle$ respectively. With $\alpha \in [0,1]$ the subspace where this Hamiltonian has non--vanishing (negative) eigenvalues is maximally two--dimensional. We represent this Hamiltonian in a basis of two orthonormal states $|v\rangle, |v^\perp\rangle$ which span this subspace. That is, we have
\ba
|w\rangle \,=\,  \beta |v\rangle + \gamma |v^\perp\rangle
\ea
where $|\beta|^2+|\gamma|^2=1$ and $\beta=\langle v| w\rangle$. The matrix elements of ${\bf H}$ restricted to this subspace are given by
\ba
{\bf H}_{| {\rm 2D}}\,=\, \begin{pmatrix} \alpha+(1-\alpha)\beta\beta^* & (1-\alpha)\beta^*\gamma \\ (1-\alpha)\beta\gamma^* & (1-\alpha)\gamma\gamma^* \end{pmatrix}  \q .
\ea
The eigenvalues for this matrix are given by
\ba
\lambda_\pm \,=\, -\frac{1}{2} \pm \sqrt{4 \alpha (\alpha-1)(1-\beta\beta^*)+1} \q .
\ea

\section{Transformation for a  tetrahedral lattice}\label{TTetra}

Here we will consider the transformation between the spin network basis and the fusion basis for a tetrahedral lattice.

The spin network basis states are given by 
\ba
{\cal W}^{j_1j_2j_3}_{j_4j_5j_6}\,=\, \pic{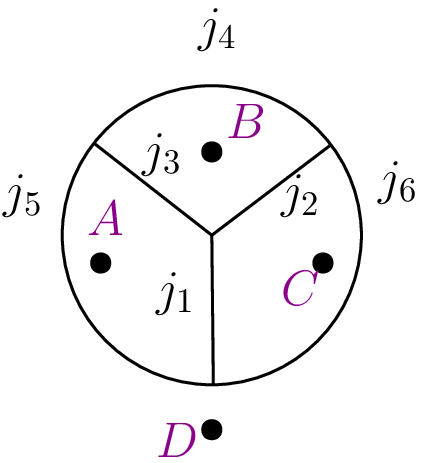}
\ea
where here we labelled  the punctures with capital latin letters.
For the fusion basis we use
\ba
{\cal O}^{k_Ak_Bk_Ck_D}_{k_{AB}k'_{AB}}\,\,=\,\, \pic{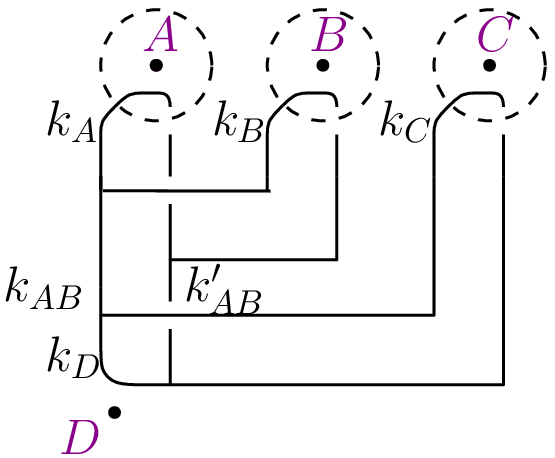} \q .
\ea


One can then find the basis transformation by computing the inner product 
between these states:
\ba
\langle  {\cal W}^{j_1j_2j_3}_{j_4j_5j_6}\, |\, {\cal O}^{k_Ak_Bk_Ck_D}_{k_{AB}k'_{AB}}\rangle
&=&
\frac{1}{{\D}^3} \sum_{l_4,l_5,l_6} \beta(j_5,k_A,l_5)\beta(j_4,k_B,l_4)\beta(j_6,k_C,l_6) \delta_{l_4l_5l_6} \nn\\
&&\bigg(\sum_{m_5} v_{j_1}v_{j_3} v_{j_4} v_{j_6}  F^{m_5 l_5 j_4}_{l_4 j_4 l_6} F^{j_5 l_5 j_5}_{j_4 j_3 m_5} F^{j_6 m_5 j_2}_{j_3 j_1 j_5} F^{j_4 l_5 m_5}_{j_6 j_2 j_6} \bigg)\nn\\
&&\bigg(\sum_{m_4 m_6}           \frac{ v_{k_A}v_{k_B}v_{k_C}   }{v_{k_{AB}}}  \, v_{m_4}  \big(R^{k_C k'_{AB}}_{m_4}\big)^* \, v_{m_6} \big(R^{k_B k_A}_{m_6}\big)^* \nn\\
&&\q\q F^{m_6 l_5 k_{AB}}_{k_Ak_B k_A} F^{k'_{AB} l_4 m_6}_{k_B k_A k_B} F^{k_D l_6 m_4}_{k_C k'_{AB}k_C} F^{l_5 l_6 l_4}_{k'_{AB} m_6 k_{AB}} F^{k_{AB} l_6 k'_{AB}}_{m_4 k_C k_D} \bigg) \, \q
\ea
where
\ba
\beta(j,k,l) &=& \frac{1}{v_k^2 v_j^2} \,\, \,\,\pic{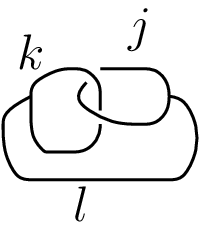} \nn\\
&=& \frac{v_l}{v_k^2 v_j^2} \sum_m v_m^2 \big( R^{jk}_m\big)^* \big( R^{jk}_m\big)^* F^{klk}_{jmj} \q .
\ea

Note that the transformation factorizes into a sum over $(l_4,l_5,l_6)$:
\ba
\langle  {\cal W}^{j_1j_2j_3}_{j_4j_5j_6}\, |\, {\cal O}^{k_Ak_Bk_Ck_D}_{k_{AB}k'_{AB}}\rangle
&=&  \sum_{l_4,l_5,l_6}    A(l_4,l_5,l_6; j_4,j_5,j_6; k_A,k_B,k_C)\,\,  \nn\\
&&\;  B(l_4,l_5,l_6; j_1,j_2,j_3,j_4,j_5,j_6)\,\, C(l_4,l_5,l_6; k_A,k_B,k_C,k_D,k_{AB},k'_{AB})\, .\q\nn\\
\ea

\end{appendix}

\end{document}